\documentclass[longauth]{aa}  
\usepackage{graphicx}
\usepackage{txfonts,array}
\usepackage{natbib}
\bibpunct{(}{)}{;}{a}{}{,} % to follow the A&A style
\usepackage[usenames,dvipsnames]{color}
\usepackage[switch]{lineno}
\usepackage{xspace}
\newcommand{\xrism}{{XRISM}\xspace}
  \newcommand{\resolve}{Resolve\xspace}
  \newcommand{\xtend}{Xtend\xspace}
\newcommand{\chandra}{\textit{Chandra}\xspace}
\newcommand{\xmm}{\textit{XMM-Newton}\xspace}
\newcommand{\nustar}{\textit{NuSTAR}\xspace}
\newcommand{\nicer}{{NICER}\xspace}
\newcommand{\swift}{\textit{Swift}\xspace}
\newcommand{\hst}{{HST}\xspace}

\begin{document} 

\title{Delving into the depths of NGC 3783 with \xrism \\ II. Cross-calibration of X-ray instruments \\ used in the large, multi-mission observational campaign }

%\author{XRISM Collaboration\thanks{Corresponding author: Jelle Kaastra}
\author{XRISM Collaboration
\and {Marc Audard}\inst{\ref{inst2}}
\and {Hisamitsu Awaki}\inst{\ref{inst3}}
\and {Ralf Ballhausen}\inst{\ref{inst4},\ref{inst5},\ref{inst6}}
\and {Aya Bamba}\inst{\ref{inst7}}
\and {Ehud Behar}\inst{\ref{inst8}}
\and {Rozenn Boissay-Malaquin}\inst{\ref{inst9},\ref{inst5},\ref{inst6}}
\and {Laura Brenneman}\inst{\ref{inst10}}
\and {Gregory V.\ Brown}\inst{\ref{inst11}}
\and {Lia Corrales}\inst{\ref{inst12}}
\and {Elisa Costantini}\inst{\ref{inst13}}
\and {Renata Cumbee}\inst{\ref{inst5}}
\and {Mar\'ia D\'iaz Trigo}\inst{\ref{inst14}}
\and {Chris Done}\inst{\ref{inst15}}
\and {Tadayasu Dotani}\inst{\ref{inst16}}
\and {Ken Ebisawa}\inst{\ref{inst16}}
\and {Megan E. Eckart}\inst{\ref{inst11}}
\and {Dominique Eckert}\inst{\ref{inst2}}
\and {Satoshi Eguchi}\inst{\ref{inst17}}
\and {Teruaki Enoto}\inst{\ref{inst18}}
\and {Yuichiro Ezoe}\inst{\ref{inst19}}
\and {Adam Foster}\inst{\ref{inst10}}
\and {Ryuichi Fujimoto}\inst{\ref{inst16}}
\and {Yutaka Fujita}\inst{\ref{inst19}}
\and {Yasushi Fukazawa}\inst{\ref{inst20}}
\and {Kotaro Fukushima}\inst{\ref{inst16}}
\and {Akihiro Furuzawa}\inst{\ref{inst21}}
\and {Luigi Gallo}\inst{\ref{inst22}}
\and {Javier A. Garc\'ia}\inst{\ref{inst5},\ref{inst23}}
\and {Liyi Gu}\inst{\ref{inst13}}
\and {Matteo Guainazzi}\inst{\ref{inst24}}
\and {Kouichi Hagino}\inst{\ref{inst7}}
\and {Kenji Hamaguchi}\inst{\ref{inst9},\ref{inst5},\ref{inst6}}
\and {Isamu Hatsukade}\inst{\ref{inst25}}
\and {Katsuhiro Hayashi}\inst{\ref{inst16}}
\and {Takayuki Hayashi}\inst{\ref{inst9},\ref{inst5},\ref{inst6}}
\and {Natalie Hell}\inst{\ref{inst11}}
\and {Edmund Hodges-Kluck}\inst{\ref{inst5}}
\and {Ann Hornschemeier}\inst{\ref{inst5}}
\and {Yuto Ichinohe}\inst{\ref{inst26}}
\and {Daiki Ishi}\inst{\ref{inst16}}
\and {Manabu Ishida}\inst{\ref{inst16}}
\and {Kumi Ishikawa}\inst{\ref{inst19}}
\and {Yoshitaka Ishisaki}\inst{\ref{inst19}}
\and {Jelle Kaastra}\inst{\ref{inst13},\ref{inst27},}\thanks{Corresponding author: Jelle Kaastra}
\and {Timothy Kallman}\inst{\ref{inst5}}
\and {Yoshiaki Kanemaru}\inst{\ref{inst16}}
\and {Erin Kara}\inst{\ref{inst28}}
\and {Satoru Katsuda}\inst{\ref{inst29}}
\and {Richard Kelley}\inst{\ref{inst5}}
\and {Caroline Kilbourne}\inst{\ref{inst5}}
\and {Shunji Kitamoto}\inst{\ref{inst30}}
\and {Shogo Kobayashi}\inst{\ref{inst31}}
\and {Takayoshi Kohmura}\inst{\ref{inst32}}
\and {Aya Kubota}\inst{\ref{inst33}}
\and {Maurice Leutenegger}\inst{\ref{inst5}}
\and {Michael Loewenstein}\inst{\ref{inst4},\ref{inst5},\ref{inst6}}
\and {Yoshitomo Maeda}\inst{\ref{inst16}}
\and {Maxim Markevitch}\inst{\ref{inst5}}
\and {Hironori Matsumoto}\inst{\ref{inst34}}
\and {Kyoko Matsushita}\inst{\ref{inst31}}
\and {Dan McCammon}\inst{\ref{inst35}}
\and {Brian McNamara}\inst{\ref{inst36}}
\and {Fran\c{c}ois Mernier}\inst{\ref{inst4},\ref{inst5},\ref{inst6}}
\and {Eric D. Miller}\inst{\ref{inst28}}
\and {Jon M. Miller}\inst{\ref{inst12}}
\and {Ikuyuki Mitsuishi}\inst{\ref{inst37}}
\and {Misaki Mizumoto}\inst{\ref{inst38}}
\and {Tsunefumi Mizuno}\inst{\ref{inst39}}
\and {Koji Mori}\inst{\ref{inst25}}
\and {Koji Mukai}\inst{\ref{inst9},\ref{inst5},\ref{inst6}}
\and {Hiroshi Murakami}\inst{\ref{inst40}}
\and {Richard Mushotzky}\inst{\ref{inst4}}
\and {Hiroshi Nakajima}\inst{\ref{inst41}}
\and {Kazuhiro Nakazawa}\inst{\ref{inst37}}
\and {Jan-Uwe Ness}\inst{\ref{inst42}}
\and {Kumiko Nobukawa}\inst{\ref{inst43}}
\and {Masayoshi Nobukawa}\inst{\ref{inst44}}
\and {Hirofumi Noda}\inst{\ref{inst45}}
\and {Hirokazu Odaka}\inst{\ref{inst34}}
\and {Shoji Ogawa}\inst{\ref{inst16}}
\and {Anna Ogorzalek}\inst{\ref{inst4},\ref{inst5},\ref{inst6}}
\and {Takashi Okajima}\inst{\ref{inst5}}
\and {Naomi Ota}\inst{\ref{inst46}}
\and {Stephane Paltani}\inst{\ref{inst2}}
\and {Robert Petre}\inst{\ref{inst5}}
\and {Paul Plucinsky}\inst{\ref{inst10}}
\and {Frederick S. Porter}\inst{\ref{inst5}}
\and {Katja Pottschmidt}\inst{\ref{inst9},\ref{inst5},\ref{inst6}}
\and {Kosuke Sato}\inst{\ref{inst47}}
\and {Toshiki Sato}\inst{\ref{inst48}}
\and {Makoto Sawada}\inst{\ref{inst30}}
\and {Hiromi Seta}\inst{\ref{inst19}}
\and {Megumi Shidatsu}\inst{\ref{inst3}}
\and {Aurora Simionescu}\inst{\ref{inst13}}
\and {Randall Smith}\inst{\ref{inst10}}
\and {Hiromasa Suzuki}\inst{\ref{inst16}}
\and {Andrew Szymkowiak}\inst{\ref{inst49}}
\and {Hiromitsu Takahashi}\inst{\ref{inst20}}
\and {Mai Takeo}\inst{\ref{inst29}}
\and {Toru Tamagawa}\inst{\ref{inst26}}
\and {Keisuke Tamura}\inst{\ref{inst9},\ref{inst5},\ref{inst6}}
\and {Takaaki Tanaka}\inst{\ref{inst50}}
\and {Atsushi Tanimoto}\inst{\ref{inst51}}
\and {Makoto Tashiro}\inst{\ref{inst29},\ref{inst16}}
\and {Yukikatsu Terada}\inst{\ref{inst29},\ref{inst16}}
\and {Yuichi Terashima}\inst{\ref{inst3}}
\and {Yohko Tsuboi}\inst{\ref{inst52}}
\and {Masahiro Tsujimoto}\inst{\ref{inst16}}
\and {Hiroshi Tsunemi}\inst{\ref{inst34}}
\and {Takeshi Tsuru}\inst{\ref{inst18}}
\and {Ay\c{s}eg\"{u}l T\"{u}mer}\inst{\ref{inst9},\ref{inst5},\ref{inst6}}
\and {Hiroyuki Uchida}\inst{\ref{inst18}}
\and {Nagomi Uchida}\inst{\ref{inst16}}
\and {Yuusuke Uchida}\inst{\ref{inst32}}
\and {Hideki Uchiyama}\inst{\ref{inst53}}
\and {Yoshihiro Ueda}\inst{\ref{inst54}}
\and {Shinichiro Uno}\inst{\ref{inst55}}
\and {Jacco Vink}\inst{\ref{inst56},\ref{inst13}}
\and {Shin Watanabe}\inst{\ref{inst16}}
\and {Brian J.\ Williams}\inst{\ref{inst5}}
\and {Satoshi Yamada}\inst{\ref{inst57}}
\and {Shinya Yamada}\inst{\ref{inst30}}
\and {Hiroya Yamaguchi}\inst{\ref{inst16}}
\and {Kazutaka Yamaoka}\inst{\ref{inst37}}
\and {Noriko Yamasaki}\inst{\ref{inst16}}
\and {Makoto Yamauchi}\inst{\ref{inst25}}
\and {Shigeo Yamauchi}\inst{\ref{inst58}}
\and {Tahir Yaqoob}\inst{\ref{inst9},\ref{inst5},\ref{inst6}}
\and {Tomokage Yoneyama}\inst{\ref{inst52}}
\and {Tessei Yoshida}\inst{\ref{inst16}}
\and {Mihoko Yukita}\inst{\ref{inst59},\ref{inst5}}
\and {Irina Zhuravleva}\inst{\ref{inst60}}
\and {Camille Diez}\inst{\ref{inst42}}
\and {Keigo Fukumura}\inst{\ref{inst61}}
\and {Chen Li}\inst{\ref{inst27},\ref{inst13}}
\and {Missagh Mehdipour}\inst{\ref{inst62}}
\and {Christos Panagiotou}\inst{\ref{inst28}}
\and {Matilde Signorini}\inst{\ref{inst24}}
\and {Keqin Zhao}\inst{\ref{inst27},\ref{inst13}}
}
\institute{
{Corresponding author: Jelle Kaastra} \and
{Department of Astronomy, University of Geneva, Versoix CH-1290, Switzerland}\label{inst2} \and %1
{Department of Physics, Ehime University, Ehime 790-8577, Japan}\label{inst3} \and %2
{Department of Astronomy, University of Maryland, College Park, MD 20742, USA}\label{inst4} \and %3
{NASA / Goddard Space Flight Center, Greenbelt, MD 20771, USA}\label{inst5} \and
{Center for Research and Exploration in Space Science and Technology, NASA / GSFC (CRESST II), Greenbelt, MD 20771, USA}\label{inst6} \and
{Department of Physics, University of Tokyo, Tokyo 113-0033, Japan}\label{inst7} \and %4
{Department of Physics, Technion, Technion City, Haifa 3200003, Israel}\label{inst8} \and %5
{Center for Space Sciences and Technology, University of Maryland, Baltimore County (UMBC), Baltimore, MD, 21250 USA}\label{inst9} \and
{Center for Astrophysics | Harvard-Smithsonian, MA 02138, USA}\label{inst10} \and %7
{Lawrence Livermore National Laboratory, CA 94550, USA}\label{inst11} \and %8
{Department of Astronomy, University of Michigan, MI 48109, USA}\label{inst12} \and %9
{SRON Netherlands Institute for Space Research, Leiden, The Netherlands}\label{inst13} \and %10
{ESO, Karl-Schwarzschild-Strasse 2, 85748, Garching bei M\"unchen, Germany}\label{inst14} \and
{Centre for Extragalactic Astronomy, Department of Physics, University of Durham, South Road, Durham DH1 3LE, UK}\label{inst15} \and %11
{Institute of Space and Astronautical Science (ISAS), Japan Aerospace Exploration Agency (JAXA), Kanagawa 252-5210, Japan}\label{inst16} \and %13
{Department of Economics, Kumamoto Gakuen University, Kumamoto 862-8680 Japan}\label{inst17} \and %15
{Department of Physics, Kyoto University, Kyoto 606-8502, Japan}\label{inst18} \and %K14
{Department of Physics, Tokyo Metropolitan University, Tokyo 192-0397, Japan}\label{inst19} \and %T16
{Department of Physics, Hiroshima University, Hiroshima 739-8526, Japan}\label{inst20} \and %17
{Department of Physics, Fujita Health University, Aichi 470-1192, Japan}\label{inst21} \and %18
{Department of Astronomy and Physics, Saint Mary's University, Nova Scotia B3H 3C3, Canada}\label{inst22} \and %19
{California Institute of Technology, Pasadena, CA 91125, USA}\label{inst23} \and
{European Space Agency (ESA), European Space Research and Technology Centre (ESTEC), 2200 AG Noordwijk, The Netherlands}\label{inst24} \and %20
{Faculty of Engineering, University of Miyazaki, 1-1 Gakuen-Kibanadai-Nishi, Miyazaki, Miyazaki 889-2192, Japan}\label{inst25} \and
{RIKEN Nishina Center, Saitama 351-0198, Japan}\label{inst26} \and %22
{Leiden Observatory, University of Leiden, P.O. Box 9513, NL-2300 RA, Leiden, The Netherlands}\label{inst27} \and %23
{Kavli Institute for Astrophysics and Space Research, Massachusetts Institute of Technology, MA 02139, USA}\label{inst28} \and %24
{Department of Physics, Saitama University, Saitama 338-8570, Japan}\label{inst29} \and %25
{Department of Physics, Rikkyo University, Tokyo 171-8501, Japan}\label{inst30} \and %26
{Faculty of Physics, Tokyo University of Science, Tokyo 162-8601, Japan}\label{inst31} \and %27
{Faculty of Science and Technology, Tokyo University of Science, Chiba 278-8510, Japan}\label{inst32} \and %28
{Department of Electronic Information Systems, Shibaura Institute of Technology, Saitama 337-8570, Japan}\label{inst33} \and %29
{Department of Earth and Space Science, Osaka University, Osaka 560-0043, Japan}\label{inst34} \and %30
{Department of Physics, University of Wisconsin, WI 53706, USA}\label{inst35} \and %31
{Department of Physics \& Astronomy, Waterloo Centre for Astrophysics, University of Waterloo, Ontario N2L 3G1, Canada}\label{inst36} \and %32
{Department of Physics, Nagoya University, Aichi 464-8602, Japan}\label{inst37} \and %33
{Science Research Education Unit, University of Teacher Education Fukuoka, Fukuoka 811-4192, Japan}\label{inst38} \and %34
{Hiroshima Astrophysical Science Center, Hiroshima University, Hiroshima 739-8526, Japan}\label{inst39} \and %35
{Department of Data Science, Tohoku Gakuin University, Miyagi 984-8588}\label{inst40} \and %36
{College of Science and Engineering, Kanto Gakuin University, Kanagawa 236-8501, Japan}\label{inst41} \and %37
{European Space Agency(ESA), European Space Astronomy Centre (ESAC), E-28692 Madrid, Spain}\label{inst42} \and %38
{Department of Science, Faculty of Science and Engineering, KINDAI University, Osaka 577-8502, JAPAN}\label{inst43} \and %39
{Department of Teacher Training and School Education, Nara University of Education, Nara 630-8528, Japan}\label{inst44} \and %40
{Astronomical Institute, Tohoku University, Miyagi 980-8578, Japan}\label{inst45} \and %41
{Department of Physics, Nara Women's University, Nara 630-8506, Japan}\label{inst46} \and %42
{Department of Astrophysics and Atmospheric Sciences,
Kyoto Sangyo University, Kyoto 603-8555, Japan}\label{inst47}
\and
{School of Science and Technology, Meiji University, Kanagawa, 214-8571, Japan}\label{inst48} \and %43
{Yale Center for Astronomy and Astrophysics, Yale University, CT 06520-8121, USA}\label{inst49} \and %44
{Department of Physics, Konan University, Hyogo 658-8501, Japan}\label{inst50} \and %45
{Graduate School of Science and Engineering, Kagoshima University, Kagoshima, 890-8580, Japan}\label{inst51} \and %46
{Department of Physics, Chuo University, Tokyo 112-8551, Japan}\label{inst52} \and %
{Faculty of Education, Shizuoka University, Shizuoka 422-8529, Japan}\label{inst53} \and %48
{Department of Astronomy, Kyoto University, Kyoto 606-8502, Japan}\label{inst54} \and %49
{Nihon Fukushi University, Shizuoka 422-8529, Japan}\label{inst55} \and %50
{Anton Pannekoek Institute, the University of Amsterdam, Postbus 942491090 GE Amsterdam, The Netherlands}\label{inst56} \and %51
{RIKEN Cluster for Pioneering Research, Saitama 351-0198, Japan}\label{inst57} \and
{Department of Physics, Faculty of Science, Nara Women's University, Nara 630-8506, Japan}\label{inst58} \and %42
{Johns Hopkins University, MD 21218, USA}\label{inst59} \and %53
{Department of Astronomy and Astrophysics, University of Chicago, 5640 S Ellis Ave, Chicago, IL 60637, USA}\label{inst60} \and  %54
{Department of Physics and Astronomy, James Madison University, Harrisonburg, VA 22807, USA}\label{inst61} \and
{Space Telescope Science Institute, 3700 San Martin Drive, Baltimore, MD 21218, USA}\label{inst62}
}

\date{\today}
 
  \abstract
   {Accurate X-ray spectroscopic measurements are fundamental for deriving basic physical parameters of the most abundant baryon components in the Universe. The plethora of X-ray observatories currently operational enables a panchromatic view of the high-energy emission of celestial sources. However, uncertainties in the energy-dependent calibration of the instrument transfer functions (e.g. the effective area, energy redistribution, or gain) can limit -- and historically, did limit -- the accuracy of X-ray spectroscopic measurements.}
   {We revised the status of the cross-calibration among the scientific payload on board four operation missions: \chandra, \nustar, \xmm, and the recently launched \xrism. \xrism carries the micro-calorimeter \resolve, which yields the best energy resolution at energies $\ge$2~keV. For this purpose, we used the data from a 10-day-long observational campaign targeting the nearby active galactic nucleus NGC~3783, carried out in July 2024.}
   {We present a novel model-independent method for assessing the cross-calibration status that is based on a multi-node spline of the spectra with the highest-resolving power (\xrism/\resolve in our campaign). We also estimated the impact of the intrinsic variability of NGC~3783 on the cross-calibration status due to the different time coverages of participating observatories and performed an empirical reassessment of the \resolve throughput at low energies.}
   {Based on this analysis, we derived a set of energy-dependent correction factors of the observed responses, enabling a statistically robust analysis of the whole spectral dataset. They will be employed in subsequent papers describing the astrophysical results of the campaign.}
   {}
    \titlerunning{X-ray cross-calibration with the NGC~3783 observational campaign}

    \authorrunning{XRISM Collaboration}
    

   \maketitle
%
%________________________________________________________________

\section{Introduction}

The launch of the X-ray Imaging and Spectroscopy Mission \citep[\xrism;][]{tashiro24} has ushered in the era of non-dispersive X-ray spectroscopy thanks to its micro-calorimeter \resolve \citep{ishisaki25, kelley25}. 
\xrism will enable unprecedented measurements of fundamental physical quantities in hot plasma, in a temperature and density regime where X-rays are the most accurate probe of its physical conditions \citep{martizzi19}.

While high-resolution spectroscopy enables diagnostic measurements of individual emission lines with little dependence on the underlying continuum or the spectral shapes at different energies, spectral analysis over broad energy ranges remains a key tool for  providing accurate estimates of plasma physical parameters, especially when observing complex absorption line systems with small equivalent widths. In this context, \xrism/\resolve spectra benefit from concurrent observations with other X-ray observatories, whose scientific payload extends the energy range and spatial coverage -- typically at the price of a moderate energy resolution and/or lower photon collecting area. In order for multi-instrument analysis to produce accurate results -- or, at least, accurate estimates of the systematic uncertainties -- a good understanding of the instrument cross-calibration is crucial. It has been shown that uncertainties in the flux and spectral shape determination from different instruments can constitute an important contribution to the uncertainty budget in some specific science cases \citep{schellenberger15}.

In this paper we present a systematic comparison between the spectra obtained by the \xrism scientific payload -- including \xtend, a charged-coupled device (CCD), large field-of-view instrument (\citealt{noda25,uchida2025}) -- and spectra obtained by \nustar \citep{harrison13}, the \xmm European Photon Imaging Camera \citep[EPIC;][]{turner01, struder01} and Reflection Grating Spectrometer \citep[RGS;][]{devries2015}, and the \chandra High-Energy Transmission Grating Spectrometer \citep[HETGS;][]{canizares05} during a multi-instrument observational campaign on the Seyfert Galaxy NGC~3783 in July 2024.
The paper is organised as follows. We describe the observational campaign in Sect.~\ref{sec:observations}. The data handling and reduction are described in Sects.~\ref{sec:datahandling} and~\ref{sec:dataReduction}. Section~\ref{sec:crosscalibration} presents the cross-calibration analysis, including a novel methodology for comparing spectral predictions from different instruments and its application to the NGC~3783 campaign. We summarise our main findings and recommendations in Sect.~\ref{sec:discussion_and_conclusions}.

\section{The NGC~3783 July 2024 observational campaign}
\label{sec:observations}

\begin{figure}
    \centering
    \includegraphics[width=1\linewidth]{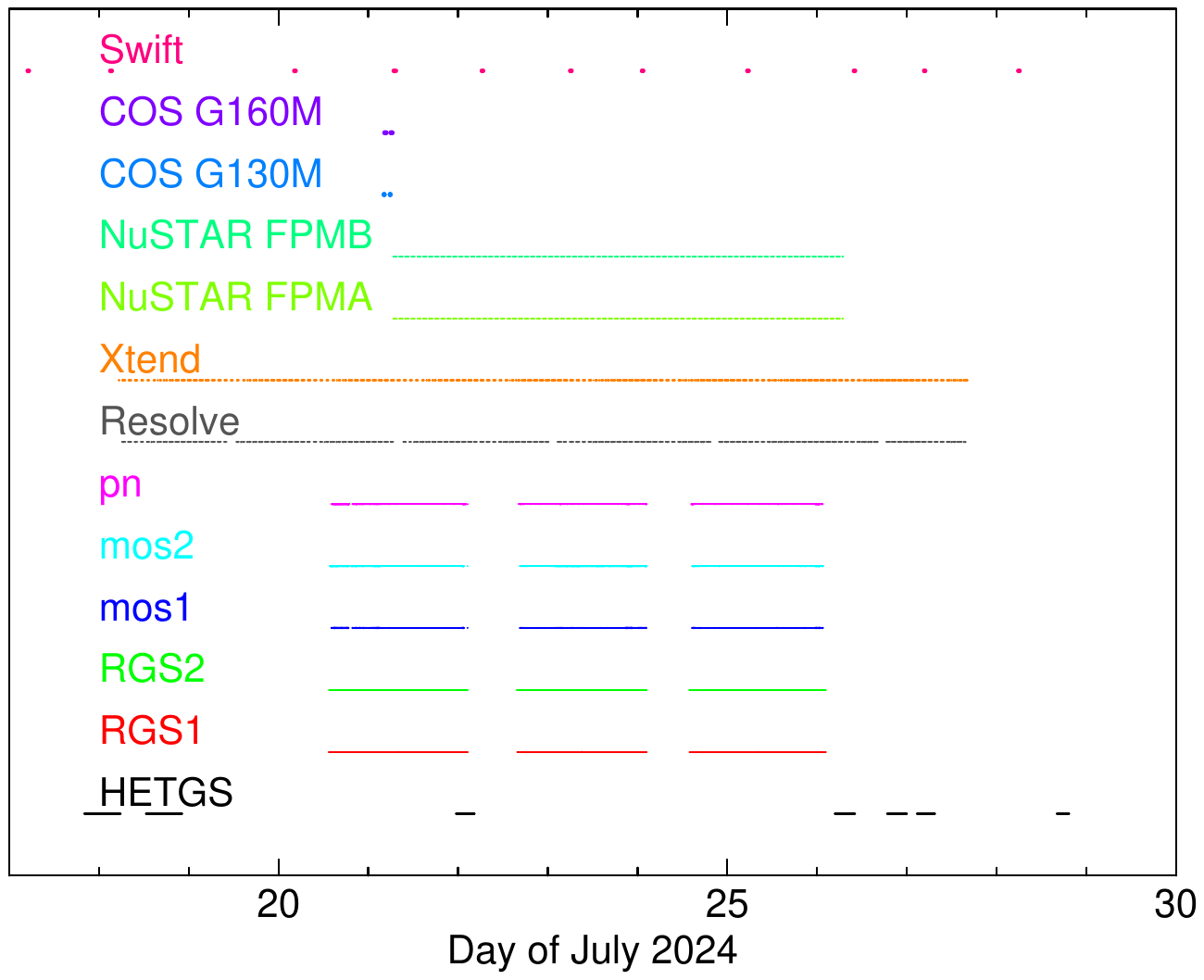}
    \caption{Overview of the observations. Horizontal lines indicate the time intervals when the instruments were observing NGC~3783. `COS' indicates the Cosmic Origin Spectrograph on board \hst. \xrism contains the \xtend and \resolve instruments, \xmm the pn, mos1, mos2, RGS1, and RGS2 instruments; for \chandra, the HETGS was the only operational instrument.}
    \label{fig:overview}
\end{figure}

The core of our campaign occurred between July 17 and 29, 2024. Coordinated observations with \chandra, the \textit{Hubble} Space Telescope (\hst), \nicer, \nustar, \swift, \xmm, and \xrism were carried out. The timeline of these observations is shown graphically in Fig.~\ref{fig:overview}. In addition to the observations shown, there were several individual \swift and \nicer snapshots preceding our main campaign, and a short (10~ks) \chandra observation in December 2023.

In the present paper we focus on the cross-calibration of the main X-ray spectroscopic instruments (\xrism, \xmm, \chandra, and \nustar). More details about the optical/UV instruments (\hst, the OM camera of \xmm, and \swift UVOT) as well as the X-ray monitoring instruments (\swift XRT and \nicer) will be given in subsequent papers where the data from these instruments are used. A summary of the observations discussed in this paper is given in Table~\ref{tab:data}.

\begin{table}[!htbp]
\caption{Observations used in this paper.}
\label{tab:data}
\centering
\begin{tabular}{lllr}
 \hline\hline
Satellite & ObsID  & Instrument   & $T_{\exp}$\tablefootmark{a}  \\
 \hline
\xrism      & 300050010            & \resolve & 439 \\
           &                      & \xtend   & 372 \\
\xmm & 0923090101,          & RGS1 & 379 \\
           & 0923090201,          & RGS2 & 381 \\
           & 0923090301           & MOS1 & 341 \\
           &                      & MOS2 & 339 \\
           &                      & pn   & 253 \\
\chandra    & 27424, 27425, 27427, & HETGS & 154 \\
           & 27366, 29487, 29488, &    & \\
           & 29495                &    & \\
\nustar     & 60901023002          & FPMA & 227 \\
\nustar     &                      & FPMB & 225 \\
\hline\noalign{\smallskip}
\end{tabular}
\tablefoot{\\
\tablefoottext{a}{Net exposure time (ks) after data screening as explained in Sect.~\ref{sec:dataReduction}.}
}
\end{table}

The X-ray spectrum of NGC~3783 is very complex and is discussed in \citet[][Paper I]{mehdipour2025} and subsequent papers. Relevant to the content of this paper, we only note a strong narrow Fe-K emission line near 6.35~keV (observed energy) and a deep Unresolved Transition Array (UTA), a blend of M-shell ions of iron between 0.7 and 0.8~keV \citep[see for example][]{mehdipour2017}.

\section{Data handling}
\label{sec:datahandling}

In this section we describe general procedures that we applied to the data of all instruments.
Data reduction details for individual instruments will be discussed in Sect.~\ref{sec:dataReduction}.
Spectral fitting was performed using SPEX software version 3.08.02 \citep{kaastra1996}. 

\subsection{Orbital corrections to the energy scale}

For the high-resolution instruments (RGS, HETGS, and \resolve), small corrections on the energy scale must be made to account for the motion of the satellite around Earth and the motion of Earth relative to the Solar System barycentre (SSB). For RGS, this is taken into account by default in the analysis software, and for HETGS and \resolve, we determined these corrections based on the ephemerid data for the satellites.

The velocity corrections $v_{\rm SSB}$ are calculated as the velocity of the satellite relative to the SSB, along the line of sight towards NGC~3783. For our campaign, these velocities are all negative, resulting in an apparent additional redshift of the spectrum (the source appears to be moving away from us).

For individual \chandra observations, the mean velocity $v_{\rm SSB}$ was in the range of $-$24 to $-$22~km/s, with an rms spread of only 0.07 km/s. It is dominated by the motion of Earth, since \chandra is most of the time far away from Earth with a low relative speed. The mean velocity is $-$22.98~km/s, and it is sufficiently accurate to apply this correction to all \chandra spectra.

For each \xrism orbit around Earth, the rms spread is 3.7~km/s, due to the higher speed of the satellite in its low-Earth orbit. For individual orbits, the average velocity ranges between $-$26 and $-$22~km/s. The mean velocity averaged over all orbits is $-$22.83~km/s.

One of the strongest absorption lines in the \chandra spectrum is the \ion{Si}{xiv} line near 2~keV. It can be measured with an accuracy of 90~km/s in the stacked spectrum. Since we typically observe about nine lines of similar strength in the full spectrum, astrophysical velocity components can be determined with accuracies of about $90/\sqrt{9}=30$~km/s. 
This justifies the use of a single and uniform velocity correction of $-$22.98~km/s for the \chandra spectra.
For \resolve, the strongest line in the spectrum is the \ion{Fe}{ii} K$\alpha$ emission line near 6.4~keV. It can be determined with a statistical uncertainty of 16~km/s in the stacked spectrum, also justifying the use of a uniform velocity correction $v_{\rm SSB}$ of $-$22.83~km/s.
Technically, we applied these velocity corrections by multiplying the model energy grid values in the response matrix by a factor of $(1+v_{\rm SSB}/c)$ where $c$ is the speed of light (note the sign!).

\subsection{Statistics}
\label{sec:statistics}

We used C-statistics instead of $\chi^2$-statistics throughout this work. The C-statistic can be used for statistical tests of the model \citep{kaastra2017} and, contrary to the $\chi^2$ statistic, it does not give a biased flux estimate \citep{hitomi2017}.

\chandra spectra are usually provided with Gehrels errors \citep{gehrels1986}, but we replaced these by the Poissonian uncertainties equal to the square root of the number of counts. This is important for both the application of the C-statistic and the optimal re-binning; re-binning Gehrels errors would lead to over-estimating the true uncertainties.

Uncertainties on parameters are reported in this paper as root-mean-square uncertainties (68\% confidence for a normal distribution).  Separate positive- and negative-side uncertainties are only provided when their absolute values differ significantly.

\subsection{Background treatment and selection}

The C-statistic has to be applied to a model for the sum of the source and background counts in the source extraction region. A model for the source counts is often based on an astrophysical model. It is not always straightforward to obtain a model for the background counts in the source extraction region. For that reason, sometimes a scaled background spectrum derived from an off-source region is subtracted. This background subtracted spectrum does not follow a Poissonian distribution, which is a fundamental requirement for the proper application of the C-statistic. Sometimes an analytical model for the background can be constructed, but in other cases the background is too complex. We constructed for all instruments except \resolve the background model from the observed background spectrum in an offset region by applying Wiener filtering to it \citep{press}. 
This reduces most of the noise in the background.  For the RGS spectra this method is applied to the individual CCDs of each spectrometer (either are active for each of the two RGS detectors). For all other instruments a single filtering procedure is sufficient. For \resolve we used a smoothed version of the non-X-ray background model instead.

The determination of the background for low-Earth orbit satellites depends on the criteria used to exclude time intervals of high-particle flux in the South Atlantic Anomaly. For \xrism we used the limits recommended in the standard data reduction pipeline for each instrument. We investigated the effect of different time screening criteria or background extraction region on the net background-subtracted \nustar spectra. Applying a stricter South Atlantic Anomaly screening to the \nustar data\footnote{\url{https://www.nustar.caltech.edu/page/background}}, the net exposure time is reduced by 6\% with only an overall $0.8\pm 1.3$\% flux difference without significant energy dependence. 

\subsection{Data binning}

For all instruments, we applied optimal binning to the spectra and response matrices \citep{kaastra2016}. The algorithm for obtaining the optimal grid for both the photon energy grid and the observed data channel grid is described in that paper. We used the new \texttt{rbin} option of the SPEX package to produce reduced size spectra and matrices. By optimally binning the data, no information is lost but the size of the matrices and thereby the time needed to convolve a model spectrum with the response matrix are strongly reduced. This is in particular important for the \resolve data, where the process reduces the size of the extra-large response matrix (see Sect.~\ref{sec:dataReduction_Resolve}) by a factor of $\sim$100 (see Table~\ref{tab:bin}).

\begin{table}[!htbp]
\caption{Number of data channels and response elements in the original and final (optimally binned) response matrix for the main instruments involved in our campaign.}
\label{tab:bin}
\centerline{
\begin{tabular}{lrrrr}
 \hline\hline
Instrument &  original & final    & original & final \\
           &  channels & channels & elements & elements \\
 \hline
 \resolve&36234 &   5694 & $1.6\times 10^8$ & $2.3\times 10^6$ \\
 \xtend  & 4096 &    241 & $8.0\times 10^6$ & $5.9\times 10^4$ \\
 FPMA   & 2005 &    248 & $6.3\times 10^6$ & $7.6\times 10^4$ \\
 RGS1   & 2855 &    671 & $8.2\times 10^6$ & $1.1\times 10^6$ \\
 RGS2   & 2767 &    671 & $7.1\times 10^6$ & $6.7\times 10^5$ \\
 MOS1   &  800 &    169 & $1.3\times 10^6$ & $1.7\times 10^4$ \\
 MOS2   &  800 &    183 & $1.3\times 10^6$ & $5.2\times 10^4$ \\
 pn     & 3329 &    162 & $9.3\times 10^5$ & $4.4\times 10^4$ \\
 MEG    & 5110 &   1704 & $1.7\times 10^6$ & $3.4\times 10^5$ \\
 HEG    & 4837 &   1451 & $5.0\times 10^5$ & $7.3\times 10^4$ \\
\hline\noalign{\smallskip}
\end{tabular}
}
\end{table}

\section{Data reduction}
\label{sec:dataReduction}

\subsection{\xrism/\resolve}
\label{sec:dataReduction_Resolve}

NGC~3783 was observed starting on July 18 , 2024, for a total \resolve duration of 451~ks (ObsID = 300050010). 
Since the \resolve aperture door (`gate valve') was closed during this observation, a $252~\mu$m-thick beryllium window attenuates X-ray photons in the soft X-ray band, reducing the effective area \citep{midooka2020}. This restricted the bandpass to energies above 1.7~keV. The aperture door also has a stainless steel protective screen that reduces the effective area by an additional  28\% across the waveband. 

The data were processed using \xrism pre-pipeline software version 004\_007.20Jun2024\_Build8.012. The data reduction was performed using \xrism pipeline processing version 03.00.012.008 and calibration database (CALDB) version 9 (20240815 release). This CALDB release includes initial updates to the \resolve energy scale and line-spread function (LSF) files based on in-orbit calibration data. The temporal gain-drift of the \resolve instrument was corrected at 6~keV using data collected during Earth occultation from $^{55}$Fe sources on the \resolve filter wheel, following the approach reported in \citet{porter25}, see also \citet{xrism2024}. 

The energy scale error at 6~keV due to intermittent gain monitoring can be estimated using the calibration pixel (pixel 12), a pixel offset from the main array that receives Mn~K$\alpha$ photons continuously from a small, collimated $^{55}$Fe source. This estimate is obtained by calculating the energy offset (`fit shift') at Mn~K$\alpha$ on the calibration pixel during the main observation period, not including the time period of the gain monitoring. This value is reported by the \xrism Software Data Center in a `\resolve Energy Scale Quality Report' for each observation.\footnote{\url{https://heasarc.gsfc.nasa.gov/docs/xrism/analysis/gainreports/index.html}} 
For the NGC~3783 observation this value is $\sim 0.16$~eV. To estimate the total uncertainty on the energy scale across the waveband, this error must be combined with the systematic energy scale errors reported in \citet{eckart25} of $\sim \pm 0.3$~eV in the 5.4--8.0~keV band, up to $\pm1.0$~eV for energies below 5.4~keV, and up to $\pm2$~eV at energies above 9 keV. The broadband energy scale errors are uncorrelated with the temporal gain-drift correction and can be root-sum-squared (see Sect. 5.3 of \citealt{eckart25}), for a total energy scale uncertainty of $\pm 0.34$~eV from 5.4--9~keV, $\pm1.0$~eV for energies below 5.4~keV, and $\pm2$~eV at energies above 9 keV. 

Additional screening and energy-dependent rise time cuts were applied according to \citet{mochizuki25}, following the procedure in the \xrism Quick Start Guide Version 2.1\footnote{\url{https://heasarc.gsfc.nasa.gov/docs/xrism/analysis/quickstart/xrism_quick_start_guide_v2p1_240506a.pdf}}. We also removed pixel 27, an edge pixel on the \resolve detector array, from our analysis. Pixel 27 has been shown to exhibit gain jumps of a few~eV on timescales of a few hours, between periods of gain monitoring, and thus current guidance recommends excluding pixel 27. After this screening, the total exposure time was slightly reduced to 439~ks. Only `high-resolution primary' events are used for the spectral analysis. 
The non-X-ray background spectrum was modelled from the \resolve night-Earth data database\footnote{https://heasarc.gsfc.nasa.gov/docs/xrism/analysis/nxb/index.html}, while the sky background was ignored because its contribution above 2 keV is negligible.

To account for detector redistribution, a redistribution matrix file was created using the {\texttt{rslmkrmf}} task with the cleaned event file and the CalDB file. This CalDB file was created using a combination of model parameters from ground-calibration measurements and updates from initial on-orbit calibration \citep{leutenegger25}. By using the extra-large option with {\texttt {rslmkrmf}}, the resulting response matrix file (RMF) incorporates all known features of the LSF model: a Gaussian core, low-energy exponential tail, escape peaks, electron-loss continuum, and Si~K fluorescence \citep{eckart18, leutenegger25}. During the NGC~3783 \resolve observation, the full width at half maximum (FWHM) of the composite Mn~K$\alpha$ spectrum of all array pixels for the calibration monitoring intervals was $4.48 \pm 0.01$~eV. The predicted FWHM resolution (core LSF) as a function of energy is provided in the CalDB file for each pixel. The uncertainty on the FWHM energy resolution (core LSF) is estimated to be 50~meV at 2~keV, 100~meV at 5~keV, and 200~meV at 8~keV for high-resolution events \citep{leutenegger25}. 
An auxiliary response file was generated using the {\texttt {xaarfgen}} task, assuming a point-like source at the aim point as input.

\subsection{\xrism/\xtend\label{sect:xtend}}

The \xtend instrument operated in 1/8-window mode to reduce frame time and mitigate potential pileup. The latest CALDB version 20241115 has been utilised in the data reduction. After applying standard event screening and filtering out periods containing high particle background, we obtained an effective exposure of 372~ks. The spectrum was extracted from the central 2\arcmin $\times$ 5.5 \arcmin\ region, while the background spectrum was obtained from an adjacent area of the same size. The background region exhibits a flat light curve, indicating no strong contamination from any solar wind events. The redistribution matrix and auxiliary response files were generated using the {\texttt {xtdrmf}} and {\texttt {xaarfgen}} routines. 

\subsection{\nustar}
\label{sec:nustar}

The \nustar data were extracted using the \nustar pipeline version 0.4.9 with the calibration database of July 29, 2024. For the spectral extraction we chose a circular region with radius 120\arcsec. For the background region we chose an annulus around the source with inner and outer radii of 180\arcsec\ and 230\arcsec, respectively, to minimise the effects of possible gradients of the background over the detector area.

\subsection{\xmm/EPIC}
\label{sec:epic}

All EPIC data were taken with the thin filter and in small window mode. EPIC source (plus background) and background spectra were extracted from calibrated and concatenated event lists. They were generated with the standard data reduction pipelines \texttt{e[mp]proc} and the latest calibration files available at the time the data reduction was performed (September 2024) using Science Analysis System (SAS) version 21.0.0 \citep{gabriel04}. Source spectra were extracted from annular regions centred around the optical active galactic nucleus (AGN) position, with an outer radius of 40\arcsec\ and 42\arcsec, and removing a core of the point spread function (PSF) corresponding to 15\arcsec\ and 20\arcsec\ for the EPIC-pn and EPIC-MOS, respectively, to mitigate mild pileup in the strongest flux states of the source during the campaign (manifesting for EPIC-pn count rates $>$19 s$^{-1}$ in the 0.3-7 keV energy band). Background spectra were extracted from circular regions of radii 45\arcsec--50\arcsec\ and 120\arcsec--140\arcsec\ for EPIC-pn and EPIC-MOS, respectively, after ensuring that the background regions do not include any serendipitous X-ray source and that the EPIC-pn background regions are at the same height in detector coordinate as the source regions to ensure that the same correction for the charge transfer inefficiency applies. Observation-specific instrument responses were generated with the standard SAS command \texttt{arfgen} and \texttt{rmfgen}.

\subsection{\xmm/RGS\label{sect:rgs}}

RGS spectra were extracted using the \xmm SAS version 21.0.0 with the following command: \texttt{rgsproc withsrc=true srcra=174.757342 srcdec=-37.738670 withrectification=yes}. This last parameter applies an empirical correction (`rectification') to the RGS effective area response aiming at aligning it to that of EPIC-pn. This correction is the default as of version 22.0.0 of SAS.

\subsection{\chandra/HETGS}

\chandra data were extracted using the standard scripts of version 4.16 of the \chandra Interactive Analysis of Observations (CIAO) package. We executed the \texttt{Chandra\_repro} process, and combined positive and negative spectral orders using \texttt{combine\_grating\_spectra}.

\section{Cross-calibration results}
\label{sec:crosscalibration}

\subsection{Effects contributing to cross-calibration uncertainty}

No instrument has a perfect calibration, and the instruments involved in our campaign are no exception. Discrepancies become apparent when the spectra from two or more instruments are compared. In general, the following factors can contribute to a mismatch between an observed spectrum and the predicted spectrum from a given astrophysical model: uncertainties in (a) the effective area, 
(b) the background modelling,
(c) pileup effects,
(d) the energy scale calibration, 
(e) the spectral redistribution function, 
and (f) time coverage. We briefly consider these effects (b--f) in the following subsections. Effective area (a) is discussed in detail in Sect.~\ref{sect:pairwise}.

\subsubsection{Background modelling}

Because NGC~3783 is a relatively bright source, the modelled background is in general small, and therefore uncertainties in the background estimate do not significantly affect the derived source spectrum. The only exception may be the higher-energy parts of the \nustar spectrum, where due to slight non-uniformity over the detector image the choice of background region has some effects (see Sect.~\ref{sec:nustar}).
Alternatively, we also extracted the background from a circular region with radius 130\arcsec\ sufficiently far away from the source region, a common practice for observers. This region yields a 1.6\% higher source flux below 30~keV and 4\% higher source flux above that energy. In our further analysis, we used the annular region centred around the source for the background.

\subsubsection{Pileup and implications of the mitigation strategy}

Pileup or count rate limits are not a problem for \xrism \resolve or for \xtend with the observing mode that we used. Only for the \xmm EPIC cameras there is some pileup in the cores of the detector image of the source (see Sect.~\ref{sec:epic}). Pileup affects the determination of the fluxes as well as the spectral shape due to pattern migration. To mitigate pileup, EPIC spectra were therefore generated from annular regions, after excising the core of the PSF primarily affected by pileup. This procedure ideally produces spectra that are fully representative of those extracted from a full aperture. However, this is not necessarily the case. Uncertainties on the encircled energy fraction calculated over the wings of the PSF may constitute the dominant -- and, regrettably, poorly known -- source of calibration uncertainties. For this reason, we postulate that the EPIC spectra of the NGC~3783 2024 observational campaign are affected by an unknown effective area calibration error. Therefore, we will not discuss the EPIC cameras in the pairwise cross-calibration analysis discussed in Sect.~\ref{sect:pairwise}.

\subsubsection{Energy-scale uncertainties}

Energy-scale uncertainties have no strong effect on high-resolution spectra, apart from a small bias in derived redshifts of lines. For lower-resolution instruments like CCDs, gain uncertainties can have a significant impact on the cross-calibration. As an example, we fitted the \xtend spectrum of NGC~3783 with a spline model (see Sect.~\ref{sec:methodology}), then applied an artificial gain shift of 2\% to the model spectrum (using the \texttt{reds} component of SPEX), and computed the residuals with respect to that redshifted model. Due to the strong Fe-K emission line, a large $\pm 10$\% oscillation of the fit residuals appears close to this line  with a significance of 18$\sigma$. It also leads to a $\sim$$5$\% large-scale deviation near 3 keV, due to the combined curvature of the spectrum and effective area of the instrument. Therefore, attention must be paid to even small uncertainties in the energy scale of moderate-resolution instruments when assessing the cross-calibration with other instruments.

\subsubsection{Spectral redistribution tails}

Uncertainties in the spectral redistribution function for high-resolution instruments do not result in major problems, because the bulk of the photons are registered in the narrow core of the redistribution function and only a small fraction ($<\sim 5$\%) appear in a near-core, low-energy tail. The situation is different for medium-resolution instruments like CCDs, in particular at low energies. Here a substantial fraction of all photons can be observed in strong, low-energy tails redistributed from the core, difficult to disentangle from photons at the bona fide correct energy. This implies that uncertainties in the low-energy effective area and the redistribution function are strongly entangled. This holds in particular for the EPIC cameras of \xmm and the \xtend detector of \xrism. 

\subsubsection{Variability and time coverage\label{sect:variability}}

Flux-dependent changes in sensitivity are not expected for our instruments. A possible complication for our analysis is the time-variability of the source. For a proper comparison, strict simultaneity is mandatory. Unfortunately, the time interval where all instruments are simultaneously operating is too small to be useful for cross-calibration purposes. Therefore, we relied on a pair-wise comparison of instruments, taking for each pair the longest strictly simultaneous time interval. The effect of time variability on some of this pair-wise analysis is discussed in Sect.~\ref{sec:sourceVariability}. For extracting spectra for a pair of instruments, their individual good time intervals (GTIs) are combined by the LHEASOFT task \texttt{ftmgtime} using only intervals where both instruments were on.

Some instruments (like the RGS on \xmm) make separate GTIs for each CCD readout sequence. When they are combined with those from another instrument that follows the same procedure, there may be no GTI left that has 100\% coverage with the other instrument, resulting in a net zero exposure time. We mitigated this by making longer GTI intervals for those instruments by first merging adjacent GTIs that are separated by less than the CCD readout sequence (a few seconds).

Another issue is related to the orbits of the low-Earth satellites \xrism and \nustar. The orbits cause interruptions in the light curve every 1.6 hour but with different phases, meaning only 30\% of the available \nustar exposure time coincides with the available \xrism exposure. Fortunately, the flux variations of NGC~3783 are not extreme. To demonstrate this, we used the power spectrum of \citet{markowitz2005} as used by \citet{li2023} to simulate the light curve. For 6500 realisations of 80 consecutive orbits (which is what we have for \nustar), the scatter in the ratio of the flux during the time \nustar was operating (50\% of the time) and the mean flux between the start and end of the \nustar observation is only 0.5\% around the mean value of 1. We conclude that for our purpose we can use safely the full \nustar dataset rather than the limited data with full overlap with \xrism.

\subsection{Methodology}
\label{sec:methodology}

Given the above-mentioned caveats, we compared the two instruments as follows.
We first performed a spline fit to the higher-resolution spectrum in the pair. For this, we used the \texttt{spln} model of SPEX, with a linear energy grid (for \resolve) or a linear wavelength grid (for the grating spectrometers).
For the resolution of the spline, we chose a number close to the resolution of the instrument, but it is optimised for any specific instrument pair.  

A great advantage of this method is that it is agnostic with respect to the complex astrophysical details of the spectrum. The best-fit C-stat for this model is often significantly smaller than the expected value for a realistic astrophysical model, because the spline also tries to accommodate for the Poissonian noise of the spectrum.
We carried out simulations using this spline method for model spectra with known shapes and found that the reconstructed spectrum with the spline model does not produce any additional bias in the source flux and thus can be used reliably.

After obtaining this best-fit model to the highest-resolution spectrum of the pair, we folded the lower-resolution instrument through this model and inspected the relative fit residuals. Re-binning the residuals by factors of typically 10 to 100 allows us to see the large-scale discrepancies of the fluxes measured by the two instruments. Based on these findings, we derived simple analytical correction factors for one instrument to get it in agreement with the other instrument, thereby making it possible to obtain broad-band spectral fits with astrophysical models by combining the two datasets. In doing that, we assumed that flux differences are solely or primarily determined by uncertainties in the calibration of the effective area. Details will be given in Sect.~\ref{sect:pairwise}.

\subsection{Global view}

\begin{figure}[!htbp]
    \centering
    \includegraphics[width=1.1\linewidth]{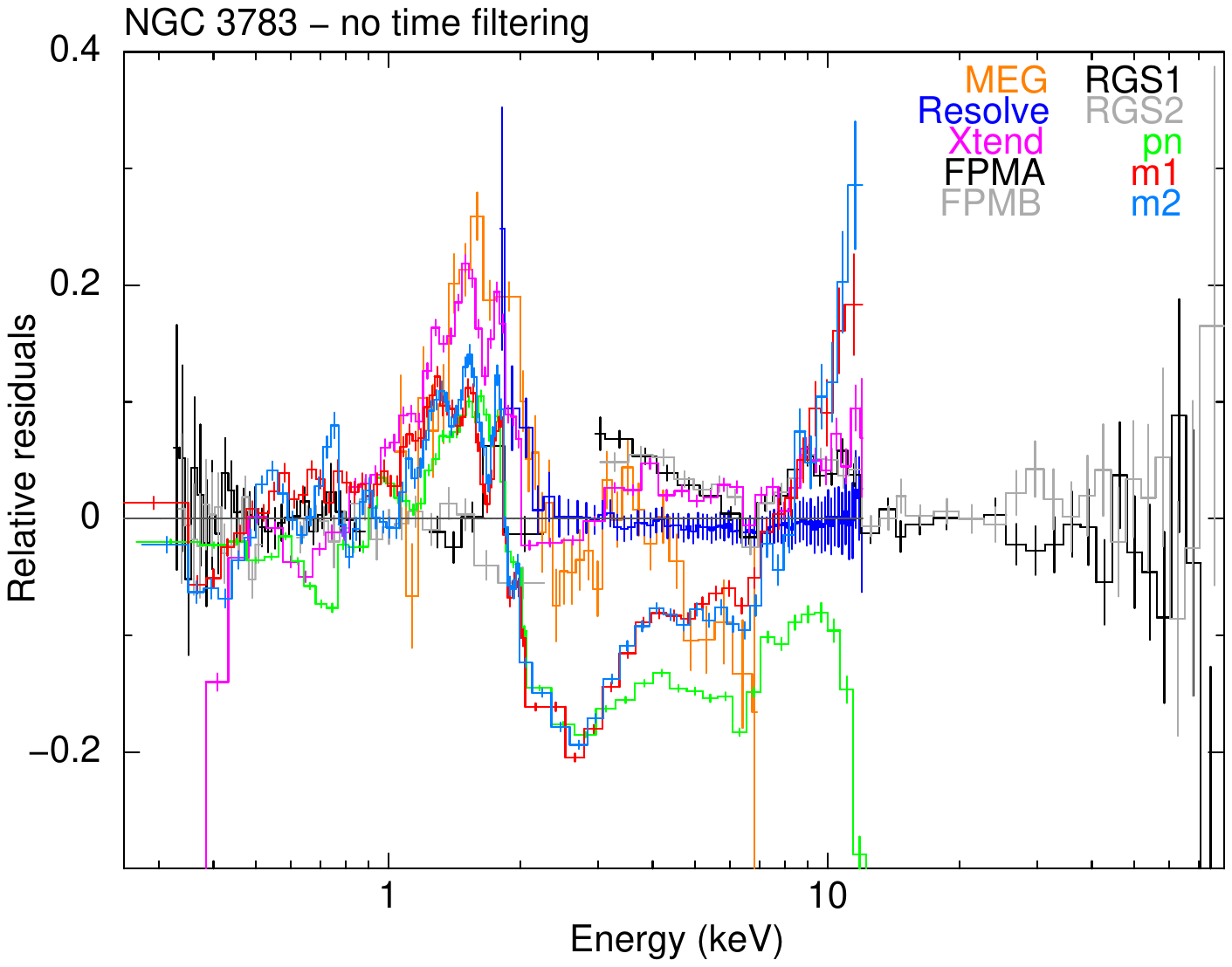}
    \caption{Relative residuals (data/model-1) against a common spline model applied to RGS, \resolve, and \nustar data (details in the text). Differences between residuals mainly indicate differences of the effective area with respect to the effective area calibrations. The residuals have been binned for clarity. The black and grey residuals refer to RGS1/2 below 3~keV and to FPMA/B above 3 keV.}
    \label{fig:global}
\end{figure}

As a first step, we made a joint fit of two high-resolution instruments (\xmm RGS and \xrism \resolve) in the band below 12~keV together with the only hard X-ray instrument (\nustar) above 12~keV using all available data. For the model, we chose three splines covering together the full energy range of these instruments, with an energy grid spacing of approximately twice the bin size of these instruments: a 326-point linear wavelength grid between 5.5 and 38~\AA\ in the RGS and lowest-energy \resolve band, a 975-point linear energy grid between 5.5~\AA\ (approximately 2.25~keV) and 12~keV in the \resolve band, and a 67-point linear energy grid between 12 and 78~keV at higher energies.
We determined the residuals for each instrument relative to this model, plus the computed residuals for the instruments that were not used for the fit: \xrism \xtend and \xmm pn, MOS1, and MOS2 (see Fig.~\ref{fig:global}).

The main differences in the relative residuals are due to calibration uncertainties; the effects of variability (different time intervals for the instruments) are only a few percent. From low to high energy, we encounter the following features. In the 1--2~keV band the RGS residuals appear to be 10-15\% below those of \chandra and \xrism. The EPIC (pn and MOS) residuals are about 20\% below those of \xrism around 3 keV, and above that energy MOS and pn start deviating strongly from each other. The \nustar fluxes in the 3--12~keV band are within a few percent of those measured by \xrism. The apparent small jump near 12~keV in the \nustar residuals is an artefact caused by using exclusively \resolve and RGS below 12~keV and \nustar above this energy.

Observers often try to compensate these kinds of residuals by adjusting the effective areas by a constant, which is a free parameter for each instruments. However, we see from Fig.~\ref{fig:global} that the residuals are strongly energy-dependent.
In order to better understand the origin of these differences and to fix them, we now turn to a detailed pair-wise comparison.

\subsection{Pair-wise instrument comparison\label{sect:pairwise}}

\subsubsection{\resolve--\xtend}

\begin{figure}
    \centering
    \includegraphics[width=1.1\linewidth]{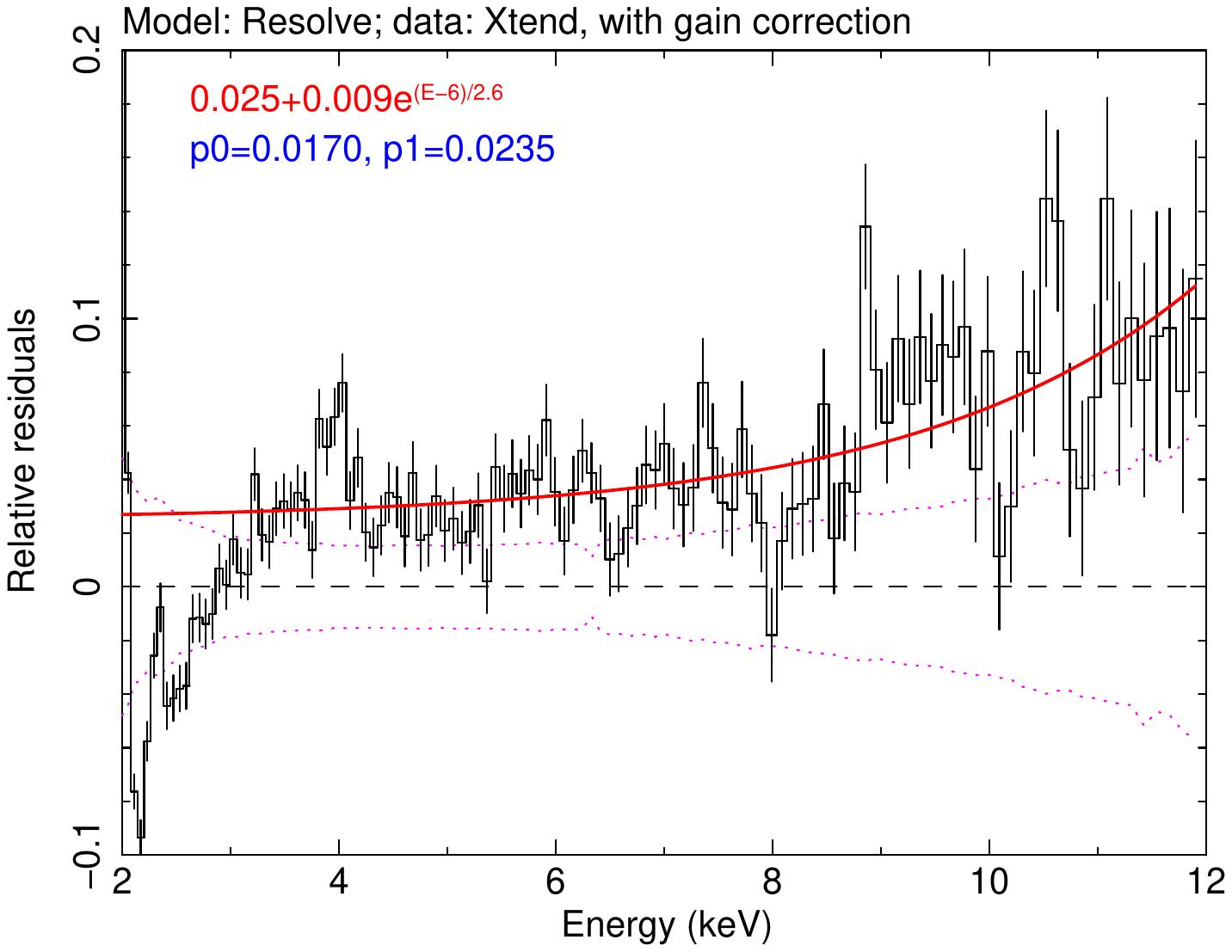}
    \caption{Residuals of the \xtend spectrum relative to the \resolve model (black). The red line is a simple analytical representation of these residuals. The dotted purple lines indicate the 1$\sigma$ combined statistical uncertainty of the \resolve spectra of NGC~3783, binned to the energy grid of the \xtend data. They represent a systematic uncertainty of the spline model for \resolve.}
    \label{fig:resolve-xtend}
\end{figure}

We produced \resolve and \xtend spectra using their common GTIs (348~ks exposure time). The \resolve spectrum was fit using a 900-point spline plus a Gaussian component for the narrow Fe-K emission line. It appears that the centroid of the Fe-K line in the \xtend spectrum is shifted by 15~eV relative to \resolve and the instrumental Si-line by 30~eV. We corrected the \xtend data here and in the rest of the paper using linear interpolation of the energy scale with the \texttt{gain} model of SPEX with two parameters $p0$ and $p1$ as indicated in Fig.~\ref{fig:resolve-xtend}. This figure shows the remaining residuals of the \xtend spectrum relative to the \resolve model. 

In the 4--8~keV energy band the two instruments agree apart from a $\sim2.5$\% offset. The narrow peak at 4~keV is not statistically significant; it is due to a $2\sigma$ dip in the \resolve model that is enhanced in \xtend due to the three times higher effective area. The deficit of 10\% near 2~keV is likely due to uncertainty in the \resolve effective area; see Sect.~\ref{sect:resolve}. There is a small excess up to $\sim 10$\% at 12~keV. This may be caused by the slightly larger size of the CCD depletion region in \xtend compared with the current response model. We approximated the excess by a simple analytical expression, as shown in the inset of Fig.~\ref{fig:resolve-xtend}.

\subsubsection{Reassessing the \resolve throughput at low energies \label{sect:resolve}}

The deficit in the \xtend spectrum near 2~keV (Fig.~\ref{fig:resolve-xtend}) can alternatively be interpreted as an excess in the flux predicted by the \resolve spectrum. A similar excess near 2~keV is observed in the \resolve data for 3C~273, which has a much simpler intrinsic spectrum in this energy range (E. Miller, private communication). Also, removal of this excess would bring the \resolve flux into closer agreement with the fluxes measured by all \xmm instruments (RGS and EPIC) at this energy.

\begin{figure}
    \centering
    \includegraphics[width=1.1\linewidth]{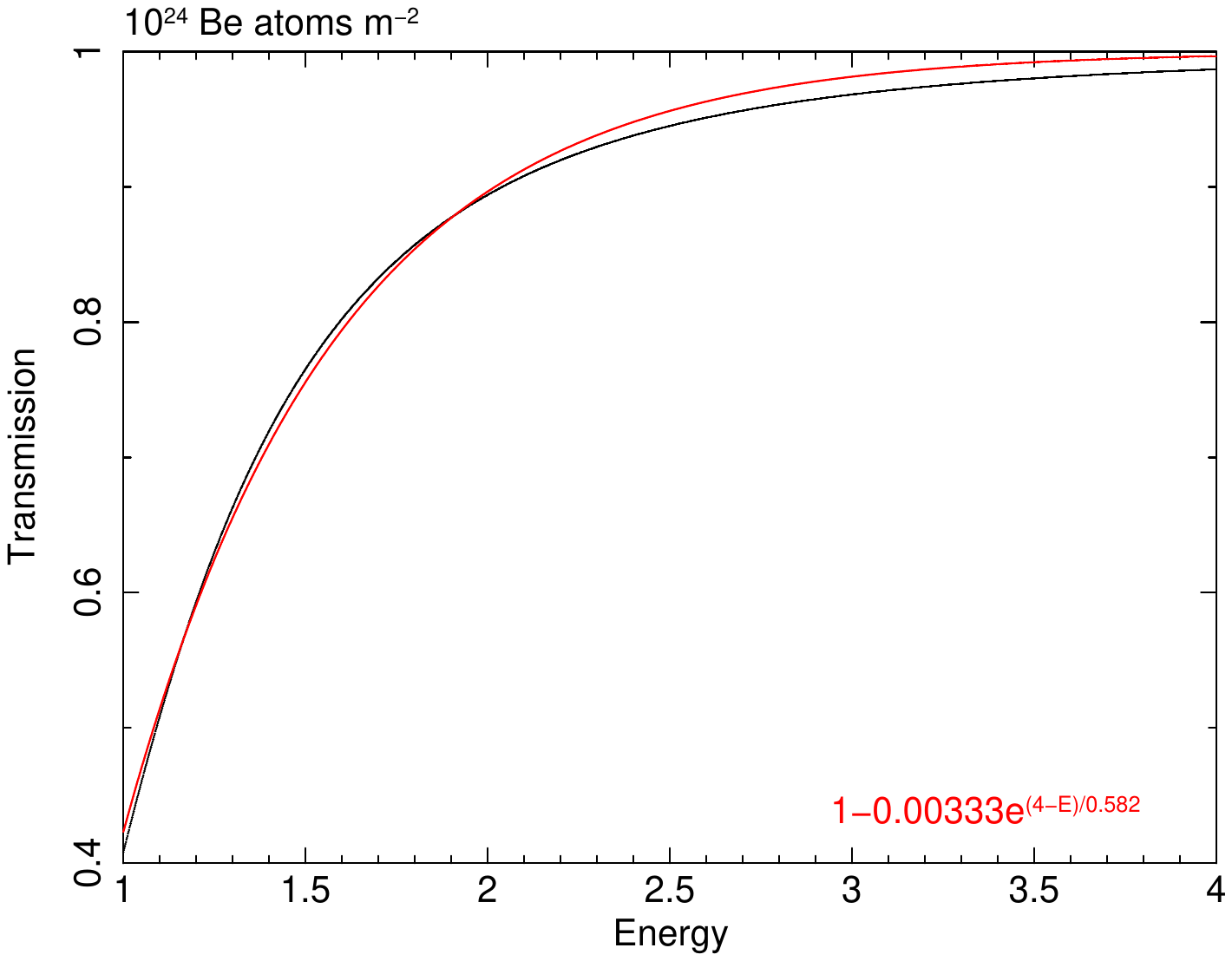}
    \caption{Simple analytical approximation (red) used to reduce excess \resolve flux compared to \xtend and other CCD instruments at the low-energy end of the \resolve band.  This expression was derived by approximating the transmission of $10^{24}$ atoms\,m$^{-2}$ (black).}
    \label{fig:betrans}
\end{figure}

We developed a simple analytical expression to adjust the \resolve flux to better match that of \xtend in this energy band. As shown in Fig.\ref{fig:betrans}, the analytical expression (red curve) approximates the transmission curve computed with SPEX (black curve). The black curve represents the transmission of a layer containing $10^{24}$ Be atoms per m$^2$, equivalent to approximately 7.5\,$\mu$m of solid beryllium. The analytical approximation was crafted for convenience, given the rapid energy-dependent variation in the transmission of the thick beryllium window of the \resolve gate valve over this waveband. This adjustment is not intended to definitively attribute the flux discrepancy between \resolve and \xtend to calibration errors or uniformity issues in the beryllium window. However, we note that the gate valve's ground calibration was performed only at energies above 2.6~keV \citep{midooka2020}, leaving the possibility that the model extrapolation to lower energies introduced error. Future measurements at a synchrotron facility using the spare unit may lead to refinement of the gate valve transmission curve at lower energies.

\subsubsection{\resolve-\nustar}

Due to the different orbital phases of \xrism and \nustar, the strictly simultaneous joint exposure time is 70~ks. We show in Sect.~\ref{sect:variability} that ignoring these orbital gaps leads to differences smaller than 0.5\%. For this reason, we compared the full \nustar spectrum (226~ks) with the full \resolve spectrum between the start and end of the \nustar observation. The \resolve spectrum was fitted using a spline between 2 and 12~keV with 801 grid points (spacing 12.5~eV) and a narrow Gaussian Fe-K line (C-stat 2475, expected 3199$\pm$80; see \citealt{kaastra2017}).

\begin{figure}
    \centering
    \includegraphics[width=1\linewidth]{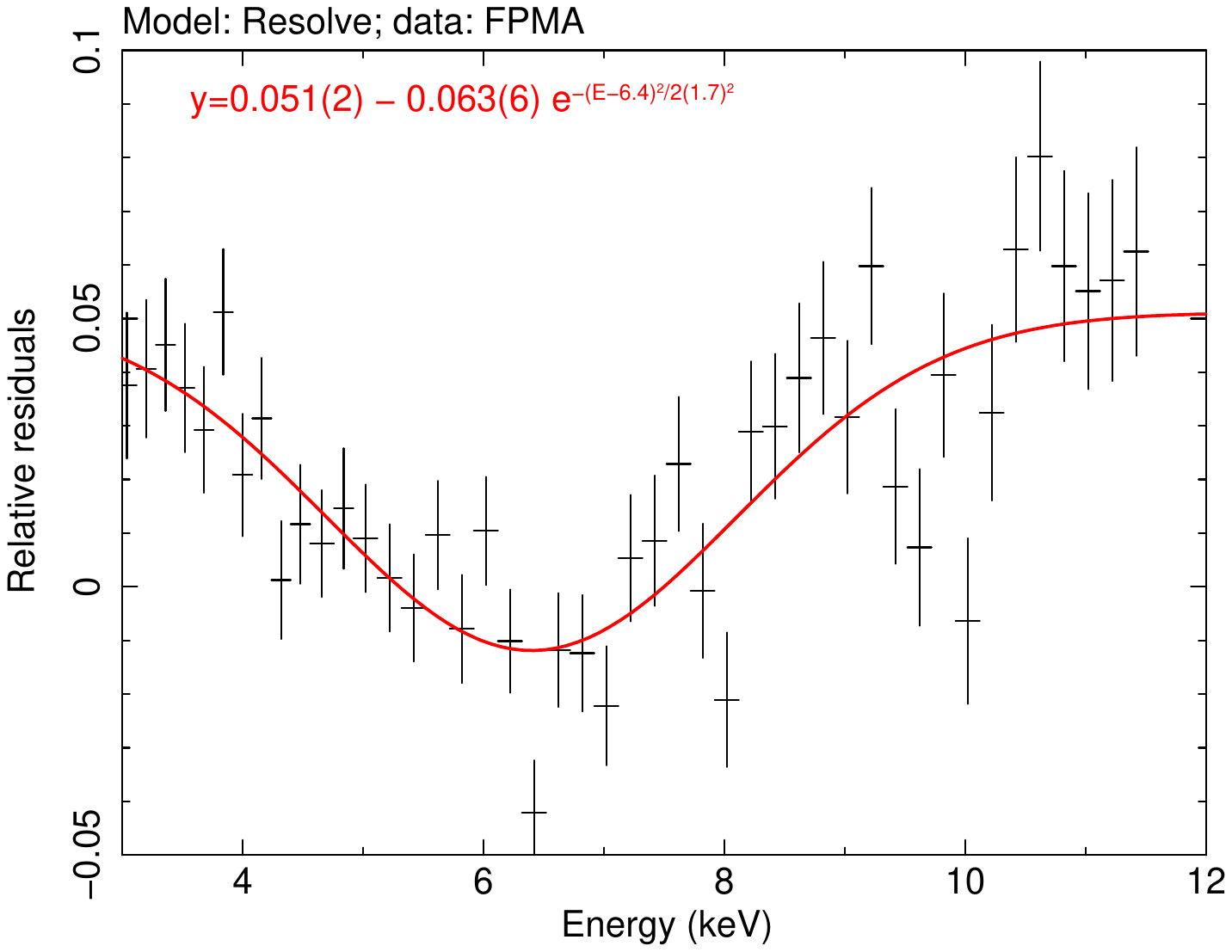}
    \caption{Residuals of the FPMA spectrum relative to the \resolve model (black). The red line is a simple analytical representation of these residuals.}
    \label{fig:resolve-fpma}
\end{figure}

Figure~\ref{fig:resolve-fpma} shows a comparison of the \nustar focal plane module A (FPMA) spectrum with this model for the \resolve spectrum. At higher energies, FPMA shows an excess of about 5\%; near 6 keV the two instruments agree, but at lower energies the residual is $+4$\%. We modelled this empirically using the formula shown in the figure.

\subsubsection{FPMA-FPMB}

We fitted a spline model to the full FPMA spectrum and investigated the residuals for FPMB. These are higher by only $0.4 \pm 0.2$\%.

\subsubsection{RGS}

Figure~\ref{fig:global} shows a strong deficit of about 15\% of the RGS and EPIC instruments of \xmm near 2 keV, relative to most other instruments. Motivated by this, we considered the effective area calibration of the RGS. The refinement of the RGS calibration is still ongoing and too complex to describe in full detail here, so we focus on an important intermediate product. 

\begin{figure}
    \centering
    \includegraphics[width=1.1\linewidth]{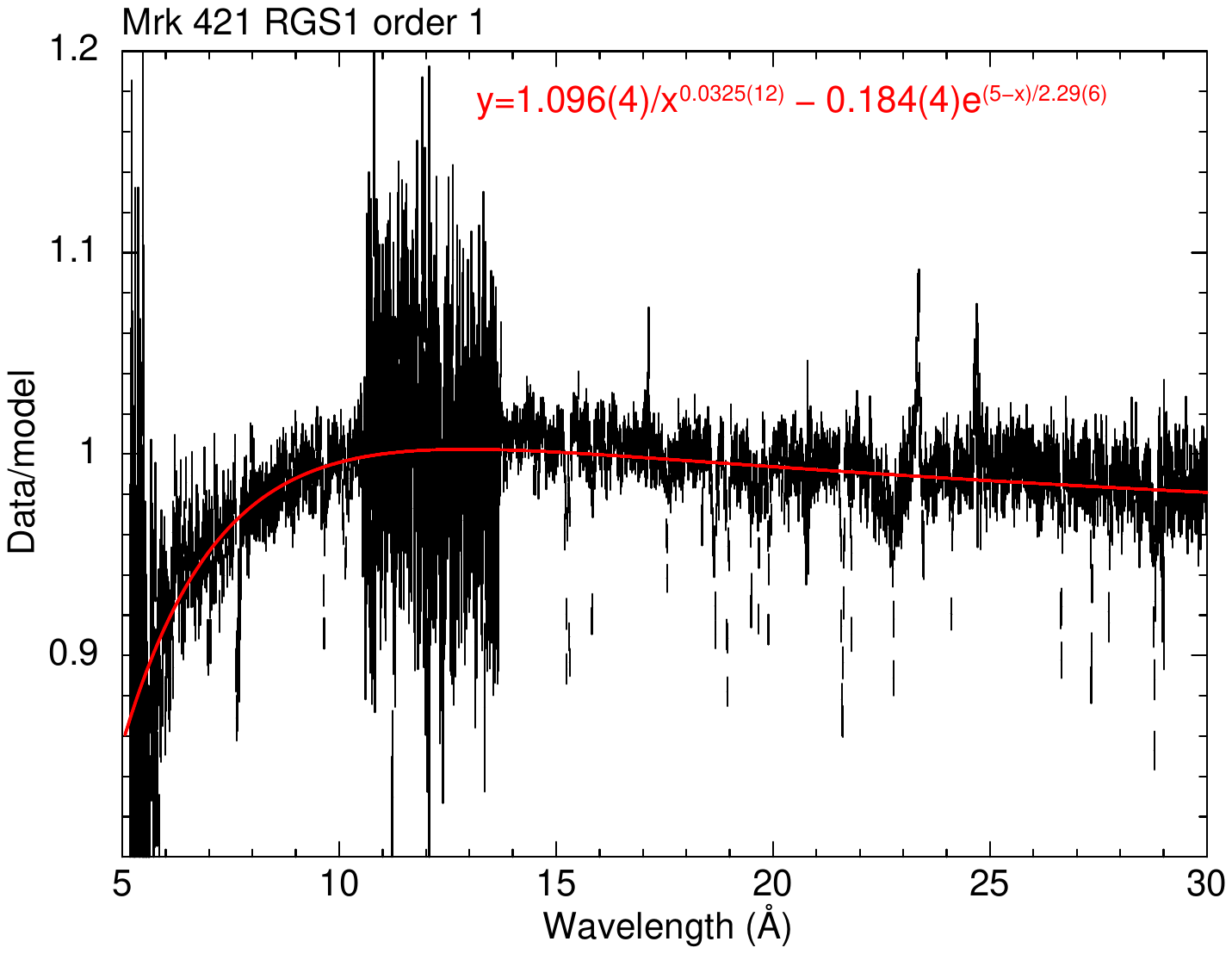}
    \includegraphics[width=1.1\linewidth]{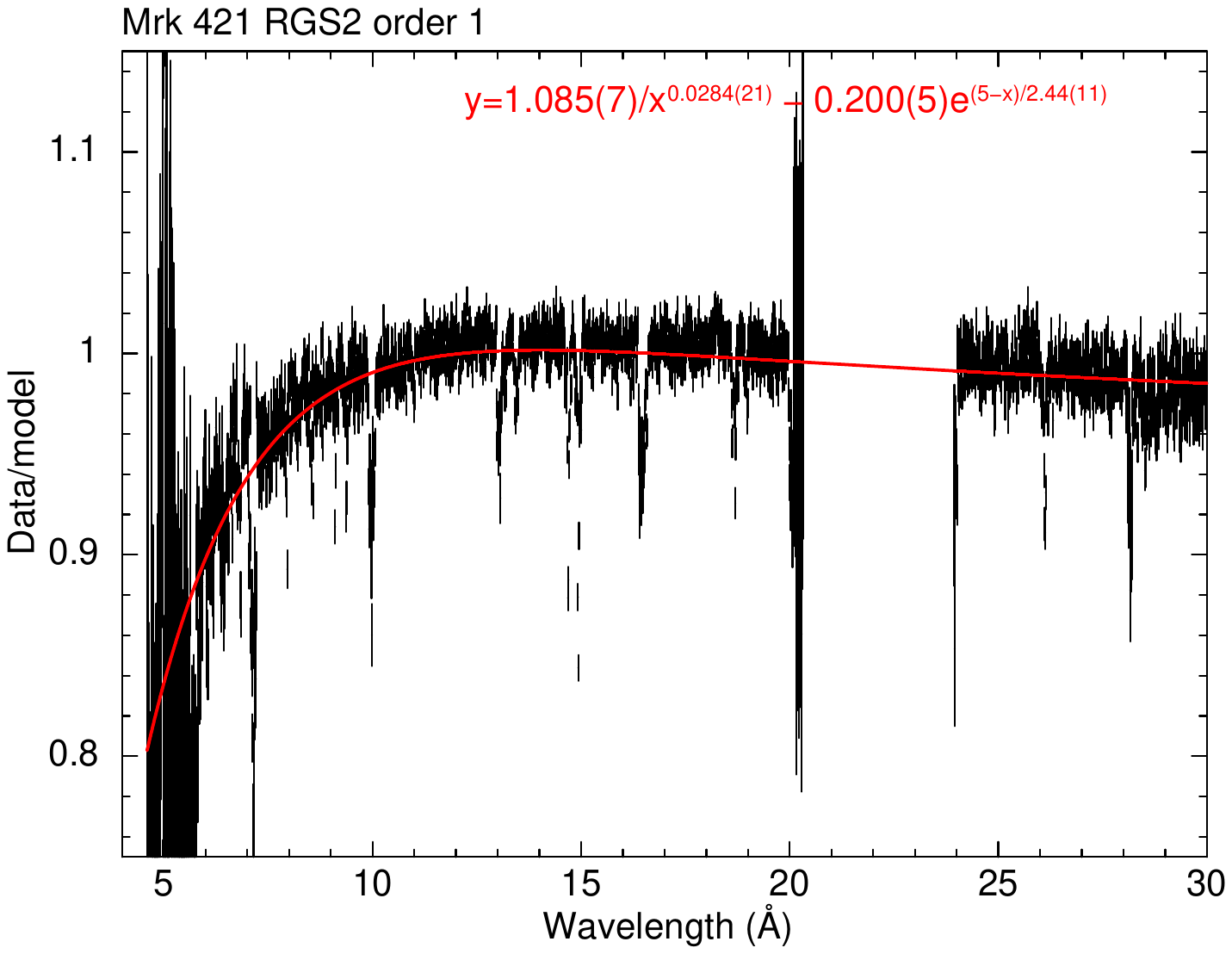}
    \caption{Stacked residuals of Mrk~421 RGS1 and RGS2 spectra as described in the text. Here $x$ represents the wavelength in \AA, and $y$ the data/model ratio. The red curves show a simple analytical model for these residuals, with shape and parameters shown in red in the insets. Negative outliers are associated with the hot or bad pixels that are not perfectly taken into account in the simplified stacking procedure.}
        \label{fig:rgs-mrk421}
\end{figure}

Figure~\ref{fig:rgs-mrk421} shows the stacked residuals of all RGS spectra collected until 2019 on the blazar Mkn~421. All individual spectra have been modelled by a broken power law with Galactic absorption. The effective area includes all default corrections except for the pn-rectification factor \citep{devries2015}. 

The figure shows that the residuals decrease rapidly for wavelengths shorter than $\sim 12$~\AA. Near 6~\AA\ or 2~keV they are about 10\% below the extrapolation of the longer-wavelength spectrum. While the cause of this decrease has not yet been explained, we made an empirical correction factor to account for it as indicated in the figures.

\begin{figure}
    \centering
    \includegraphics[width=1.1\linewidth]{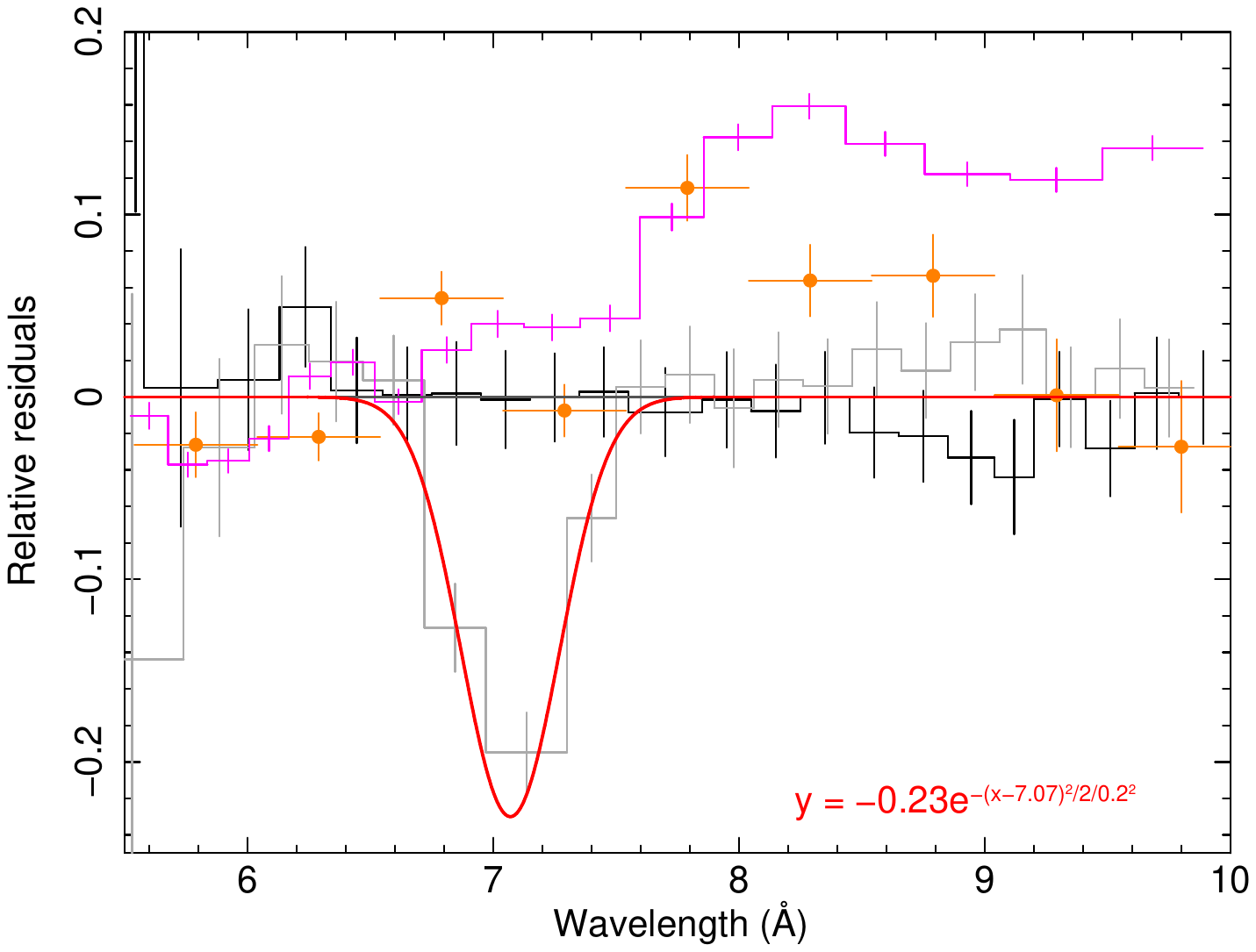}
    \caption{Residuals for a joint fit with a spline model through the RGS1 (black), RGS2 (grey, excluding wavelength between 6.55 and 7.55~\AA\ in the fit), \resolve and \nustar (not shown) instruments. For comparison we show the residuals to this model for \xtend (purple histogram) and MEG (orange); these last two instruments were not included in the fit. The red line is a fit to the RGS2 residuals, with model and parameters indicated in the inset.}
    \label{fig:rgs2fixed}
\end{figure}

With this correction taken into account, we made a new joint spline fit through the RGS, \resolve, and \nustar spectra, taking into account the effective area corrections for these instruments obtained so far. These results will be discussed in more detail later, but they show an additional discrepancy between RGS1 and RGS2 around 7~\AA\ (Fig.~\ref{fig:rgs2fixed}). As discussed below, this discrepancy can be attributed to RGS2. Therefore, we excluded RGS2 between 6.55 and 7.55~\AA\ in the fit.

Figure~\ref{fig:rgs2fixed} shows that around 7~\AA, RGS2 deviates up to $\sim$23\% from RGS1, \xtend and HETGS medium-energy grating (MEG). A broad-band fit with a realistic astrophysical model through the combined \resolve and RGS spectra (Zhao et al., in prep.) confirms that RGS1 represents the source spectrum properly near 7~\AA. The RGS calibration becomes more uncertain for wavelengths below 7~\AA, and in particular near chip gaps. Such a gap occurs for RGS2 between CCDs 1 and 2 near 7.2~\AA. For this reason, we applied the empirical correction shown in Fig.~\ref{fig:rgs2fixed} to the RGS2 data.

\subsubsection{RGS-\xtend}

\begin{figure}
    \centering
    \includegraphics[width=1.1\linewidth]{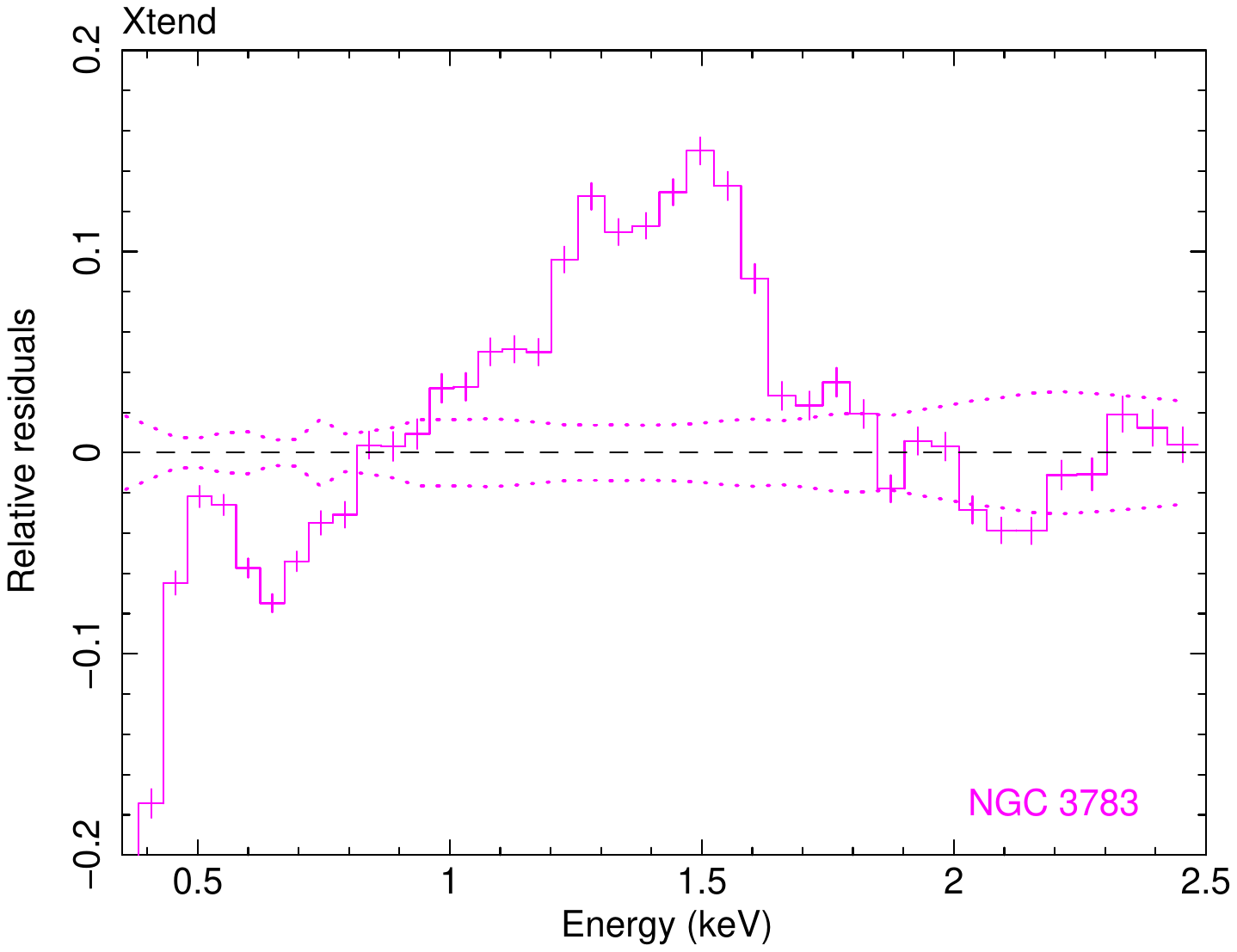}
    \caption{Relative residuals of the \xtend spectrum of NGC~3783 relative to a spline fit to the simultaneous RGS, \resolve, and \nustar spectra. The dotted purple lines indicate the 1$\sigma$ combined statistical uncertainty of the combined RGS+\resolve spectra of NGC~3783, binned to the energy grid of the \xtend data. They represent a systematic uncertainty of the spline model for NGC~3783.}
    \label{fig:3c273}
\end{figure}

After incorporating the corrections to the RGS, \resolve and \nustar spectra presented above, we folded the \xtend spectrum through this model (Fig.~\ref{fig:3c273}). 
In addition, we made a spline fit to the RGS spectrum of another radio-loud active AGN, 3C273. This source is observed every year in the framework of an international cross-calibration campaign involving all operational X-ray observatories. The RGS and \xtend data were reduced as specified in Sects.~\ref{sect:rgs} and~\ref{sect:xtend}, respectively. We folded that model through the simultaneous \xtend data of that source, which were analysed in the same way as the NGC~3783 data. At least for energies larger than about 0.8~keV, the residuals are similar to those shown in Fig.~\ref{fig:3c273} and therefore point to a common discrepancy between the effective area of RGS versus \xtend. A detailed analysis of the 3C~273 data will be presented in a separate paper (Miller et al., in prep.).

\subsubsection{\resolve-\chandra HETGS}

\begin{figure}
    \centering
    \includegraphics[width=1.1\linewidth]{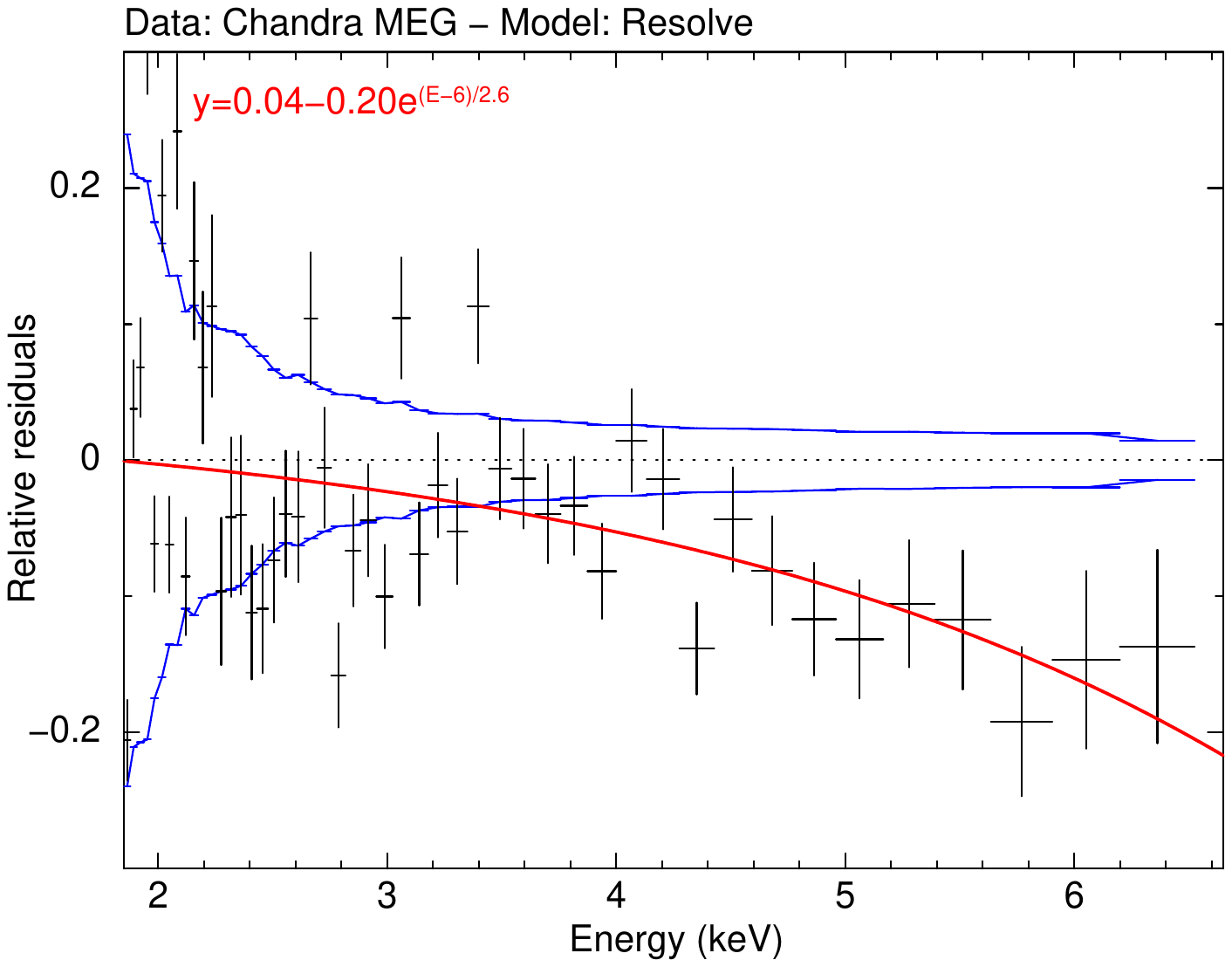}
    \caption{Relative residuals of the \chandra HETGS-MEG spectrum (binned by a factor of 10) relative to the \resolve model (black). The red line is a simple analytical representation of these residuals. Blue lines indicate the $\pm 1\sigma$ uncertainty on the \resolve spectrum binned to the MEG bins.}
    \label{fig:resolve-meg}
\end{figure}

We fitted the \resolve spectrum between 1.8 and 6.7 keV using a spline with 981 knots (spacing 5~eV). The \resolve spectrum was modified according to the empirical correction of the soft X-ray throughput described in Sect.~\ref{sect:resolve}, which affects the low-energy part of the spectrum. Due to the small effective area at low energies the fluxed \resolve spectrum becomes uncertain, as shown by the blue lines in Fig.~\ref{fig:resolve-meg}.

\subsubsection{\chandra HEG and MEG}

\begin{figure}
    \centering
    \includegraphics[width=1.1\linewidth]{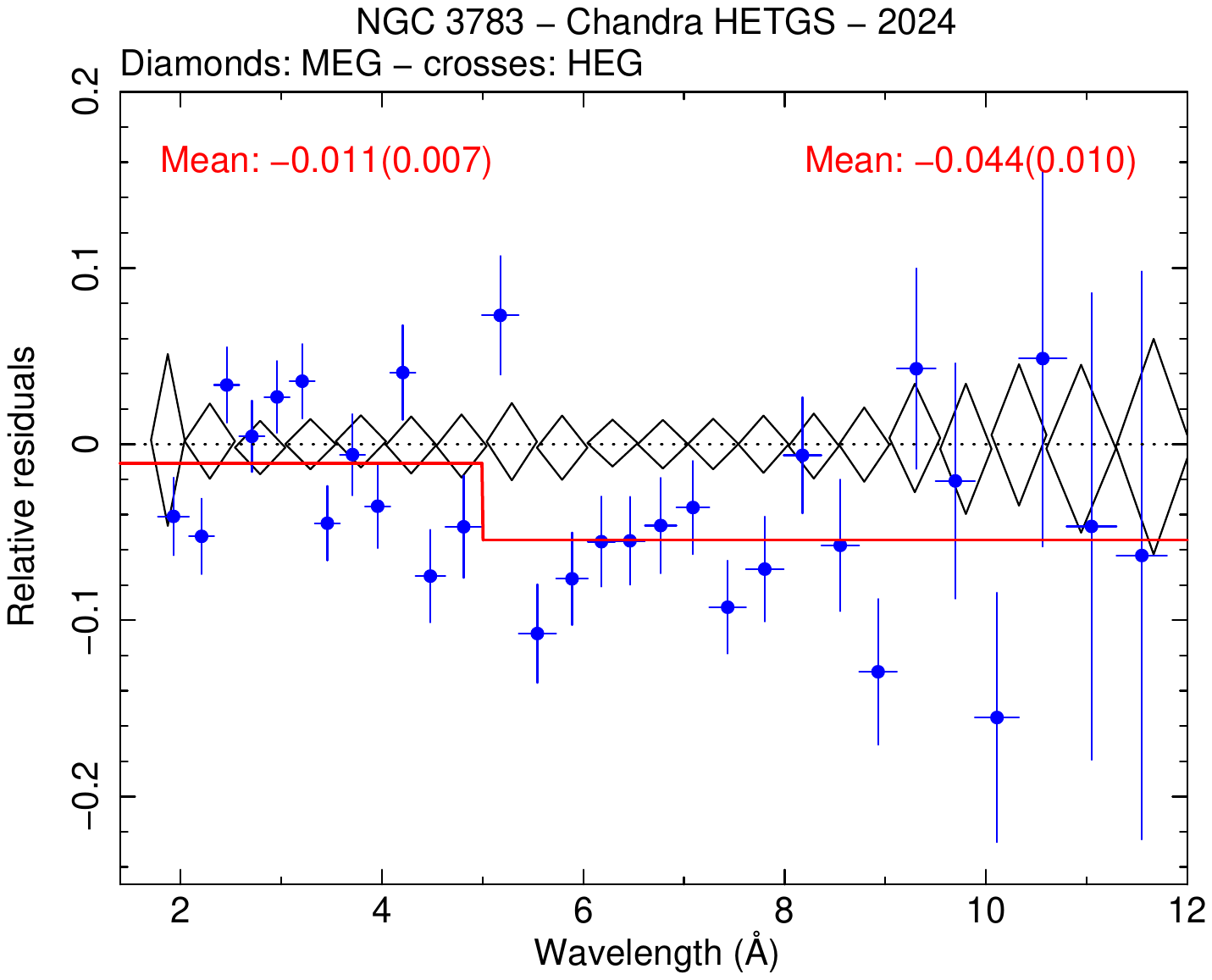}
    \caption{Relative residuals of a spline fit to the \chandra MEG spectrum, and the HEG spectrum relative to that model.}
    \label{fig:chandra}
\end{figure}

A comparison of the two gratings of the HETGS can be made using the full dataset, because the two operate simultaneously. We fitted the MEG spectrum with a spline and folded the high-energy grating (HEG) spectrum through it. The ratio is $\simeq -1$\% for $\lambda\le -5$~\AA, and $\simeq -4$\% for $\lambda\ge -5$\AA\ (Fig.~\ref{fig:chandra}).

\subsection{Corrections for source variability for joint instrument analysis}
\label{sec:sourceVariability}

The ultimate goal of this paper requires production of spectra that can be analysed together. In particular when the time-averaged spectrum is considered, corrections must be made for the different time intervals when the instruments were operating. Fortunately, the time interval during which \xtend was operated encompasses the operational time intervals of the other instruments. Therefore, for each instrument, we can determine the ratio of the 0.3--12~keV band \xtend flux when that other instrument was operating, against the flux measured during the full \xtend exposure. The major contribution to this ratio is due to the variability of the primary AGN power-law component. Variations on smaller energy scales cannot be fully excluded using the medium-resolution \xtend data, and would need special care in variability studies using the high-resolution instruments involved, but we expect their impact to be small on the X-ray broadband continuum.

\begin{figure}
    \centering
    \includegraphics[width=1.1\linewidth]{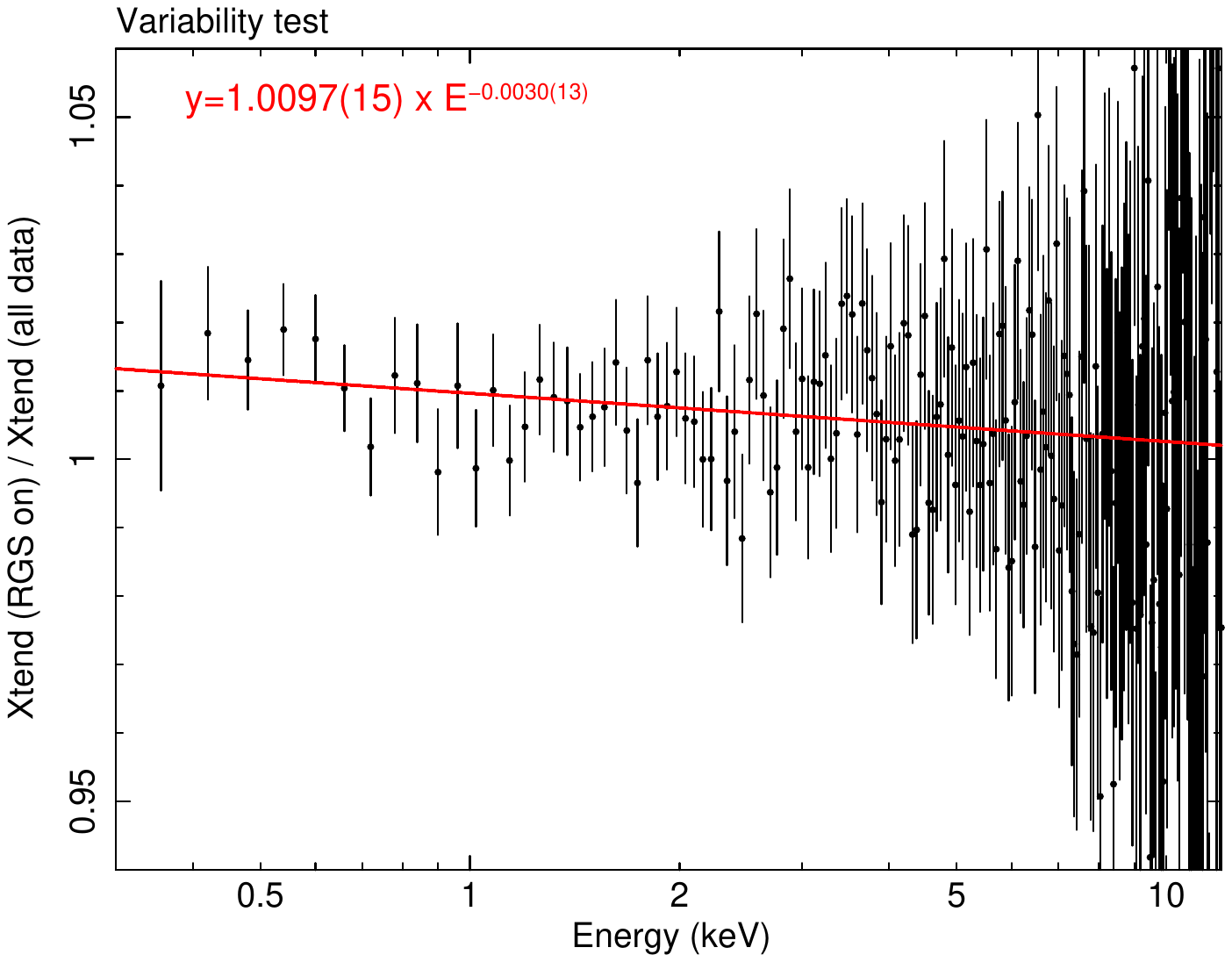}
    \caption{Ratio of the \xtend spectrum during the time when RGS was observing to the full \xtend spectrum. The red curve is a simple power-law approximation to this ratio.}
    \label{fig:rgsvar}
\end{figure}

Figure~\ref{fig:rgsvar} shows the ratio $y$ of the \xtend spectra between the full exposure and the common time interval when RGS was also being operated. It can be approximated by a simple power law. Similar results were found for the \chandra and \nustar data. Best-fit power-law parameters $A$ and $b$ ($y=AE^b$ with E in keV) are given in Table~\ref{tab:variability}. The deviations from unity are significant but small, except for the \chandra data, and will be taken into account in the analysis of the combined dataset in future papers. 

\begin{table}[!htbp]
\caption{Parameters $A$ and $b$ for power-law fits to the ratio of \xtend spectra extracted during a time interval common to another instrument and during the full exposure.}
\label{tab:variability}
\centerline{
\begin{tabular}{lrrrr}
 \hline\hline
Instrument &  A & b  & $f_1$ & $f_2$   \\
 &  &  \\
 \hline
 RGS    & 1.0097(15) & -0.0030(13) & 1.012 & 1.002 \\
 \nustar & 1.0226(14) & -0.0043(13) & 1.027 & 1.012 \\
 \chandra& 1.113(4)   &   -0.017(3) & 1.130 & 1.068 \\
\hline\noalign{\smallskip}
\end{tabular}
}\tablefoot{The values of the best-fit power-law function $f_1$ and $f_2$ correspond to the lowest ($E_1=0.3$~keV) and highest ($E_2=12.5$~keV) boundary of the \xtend energy bandpass.}
\end{table}

For \chandra, five of the seven observations have been taken while \xrism was also observing NGC~3783. The ratio of this \chandra spectrum to the full \chandra spectrum can be approximated by $1+0.07/\lambda^{1.2}$ with $\lambda$ the wavelength of a photon in \AA.

\section{Discussion and conclusions}
\label{sec:discussion_and_conclusions}

Based on our partially simultaneous data from multiple instruments and considerations of the cross-calibration of these instruments, we plan to make adjustments to selected instrument effective area curves in our subsequent papers, which will discuss multi-instrument data taken during the NGC~3783 2024 observational campaign. Relative to the standard calibration available at the time of this paper, corrections will be applied to the low- or high-energy side of the effective area curves, as summarised in Table~\ref{tab:calsummary}. The factor $f(E)$ listed in the table represents the multiplier to the effective area. 

\begin{table}[!htbp]
\caption{Energy-dependent multiplication factors to be applied to the effective area for different instruments. }
\label{tab:calsummary}
\setlength\extrarowheight{3.5pt}
\centerline{
\begin{tabular}{ll}
 \hline\hline
Instrument &  $f$   \\
 \hline
 \resolve & $f=1 / (1-0.00333 {\rm e}^{(4-E)/0.582} )$ \\
 \xtend & $f=1.025 + 0.009{\rm e}^{(E-6)/2.6}$\\
 RGS1  & $f= 1.096/\lambda^{0.0325} -0.184{\rm e}^{(5-\lambda)/2.29}$\\
 RGS2  & $f= 1.085/\lambda^{0.0284} -0.200{\rm e}^{(5-\lambda)/2.44}$\\
 & $- 0.23{\rm e}^{0.5(\lambda-7.07)^2/0.20^2}$ \\
 \nustar & $f=1.051 -0.063{\rm e}^{0.5(E-6)^2/1.7^2} $\\
 \chandra MEG &  $f=1.04-0.20{\rm e}^{(E-6)/2.6}$\\
\hline\noalign{\smallskip}
\end{tabular}
}\tablefoot{In these equations, $E$ is the energy in keV and $\lambda$ the wavelength in \AA.}
\end{table}

Because the data are not strictly simultaneous, additional corrections need to be made to compensate for missing data. This is facilitated by the fact that AGNs like NGC~3783 often show a strong correlation between the photon index and flux of the dominant power-law component, at least in the 2--10~keV band. We determined these correction factors by comparing the full \xtend spectra with \xtend spectra during the time that the other instrument was observing NGC~3783. The correction factors are well represented by power-law corrections (Table~\ref{tab:variability}).

\begin{figure}
    \centering
    \includegraphics[width=1.1\linewidth]{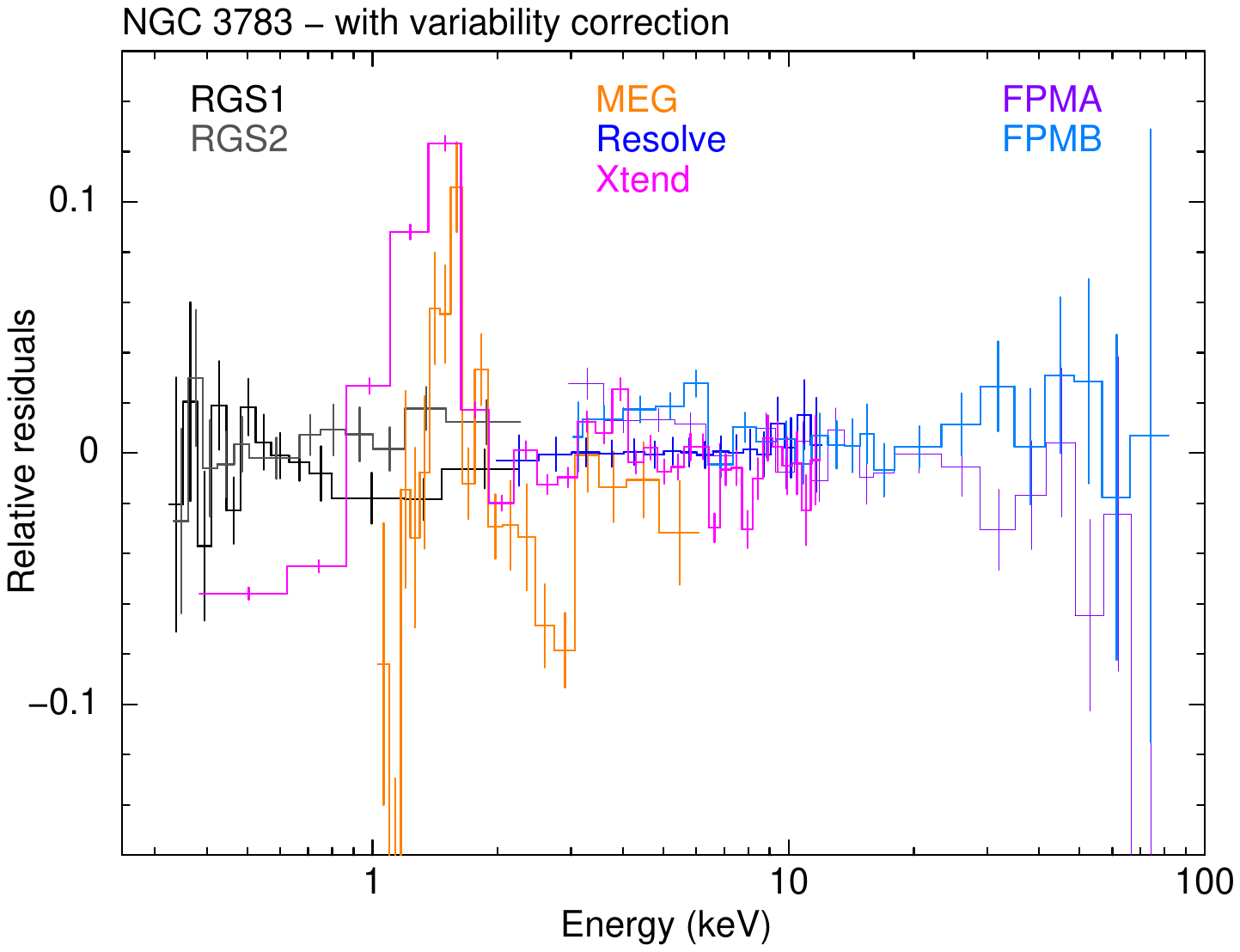}
    \caption{Relative residuals of the spectra. The residuals of RGS and MEG have been binned by a factor of 50, for \resolve a factor of 200, \nustar a factor of 5 below 18 keV and 20 above 18 keV, and a factor of 5 for \xtend.}
    \label{fig:calcheck}
\end{figure}

We verified these correction factors by making a spline fit to the combined RGS, \resolve, and \nustar data (the \nustar data were restricted to the band above 10~keV in the fit). We show the residuals relative to this model for the full bands of these instruments and for \xtend and the \chandra MEG in Fig.~\ref{fig:calcheck}.
While the adjusted \resolve, \nustar, and RGS spectra agree within 1--2\% with this model, showing the effectiveness of our procedure, the \chandra MEG spectrum still shows some fluctuations up to $\sim\pm$6\% in the 1--3~keV band. This is likely due to the relatively small overlap between the \xrism and \chandra data. As previously noted, two of the seven \chandra spectra were taken before or after the \xrism observations, and the other five overlap with only about 10\% of the duration of the \xrism observation. In fact, most of the overlapping \chandra exposure is near the start or the end of the \xrism observation, epochs characterised by relatively high flux. Thus, the discrepancies may be caused by spectral variations in the astrophysical source. This will be explored in forthcoming papers.

The \xtend spectra below 2~keV still show differences of up to 14\% relative to the model, which is dominated here by the RGS instrument. This may be related to the complexity of the calibration of CCD detectors at low energies, or the effective area calibration of the RGS, but more effort, beyond the scope of this paper, is needed to test that hypothesis.

We plan to apply the empirical corrections described in Tables~\ref{tab:variability} and~\ref{tab:calsummary} to subsequent papers, which will present the multi-instrument analysis of the observational campaign of NGC~3783. These corrections are not based on a physical model of the telescope or the instruments. We therefore discourage the reader from interpreting the results in this paper as `the cross-calibration status' between the \xrism scientific payload and other operational X-ray observatories. Effective area calibration may subtly depend on source properties and observational configuration. A full understanding of the typical differences in the determination of flux and spectral shape between \xrism and other observatories can only be derived from an analysis of the full sample of sources included in the \xrism in-flight calibration plan \citep{miller20}. Furthermore, our understanding of the \xrism payload calibration is still in its initial stages. Changes in the effective area of the \resolve and \xtend instruments are likely to be encoded in future versions of the calibration database. The status we describe in this paper -- and in particular through Fig.~\ref{fig:global} -- constitutes an encouragement to carefully take cross-calibration uncertainties into account  before undertaking multi-instrument global fits involving the \xrism scientific instruments. 

\begin{acknowledgements}

We are grateful to the mission planners of all missions involved for their efforts in scheduling and re-scheduling all our coordinated observations as well as possible, which is not a trivial task.
We thank Lucien Kuiper (SRON) for providing us with the velocity corrections to the Solar System Barycentre. SRON is supported financially by NWO, the Netherlands organization for scientific research. 

Part of this work was performed under the auspices of the U.S. Department of Energy by Lawrence Livermore National Laboratory under Contract DE-AC52-07NA27344. The material is based upon work supported by NASA under award number 80GSFC21M0002. This work was supported by JSPS KAKENHI grant numbers JP22H00158, JP22H01268, JP22K03624, JP23H04899, JP21K13963, JP24K00638, JP24K17105, JP21K13958, JP21H01095, JP23K20850, JP24H00253, JP21K03615, JP24K00677, JP20K14491, JP23H00151, JP19K21884, JP20H01947, JP20KK0071, JP23K20239, JP24K00672, JP24K17104, JP24K17093, JP20K04009, JP21H04493, JP20H01946, JP23K13154, JP19K14762, JP20H05857, and JP23K03459. This work was supported by NASA grant numbers 80NSSC20K0733, 80NSSC18K0978, 80NSSC20K0883, 80NSSC20K0737, 80NSSC24K0678, 80NSSC18K1684, and 80NNSC22K1922. LC acknowledges support from NSF award 2205918. CD acknowledges support from STFC through grant ST/T000244/1. L. Gallo acknowledges financial support from Canadian Space Agency grant 18XARMSTMA. MS acknowledges the support by the RIKEN Pioneering Project Evolution of Matter in the Universe (r-EMU) and Rikkyo University Special Fund for Research (Rikkyo SFR). AT and the present research are in part supported by the Kagoshima University postdoctoral research program (KU-DREAM). SY acknowledges support by the RIKEN SPDR Program. IZ acknowledges partial support from the Alfred P. Sloan Foundation through the Sloan Research Fellowship. This work was supported by the JSPS Core-to-Core Program, JPJSCCA20220002. The material is based on work supported by the Strategic Research Center of Saitama University.
M. Mehdipour acknowledges support from NASA through the XRISM Guest Scientist (XGS) grant 80NSSC23K0995, a grant for HST program number 17273 from the Space Telescope Science Institute (operated by the Association of Universities for Research in Astronomy, Inc., under NASA contract NAS5-26555), and NuSTAR grant 80NSSC25K7126.

We deeply regret that our co-author Katja Pottschmidt passed away on 17 June 2025, the same day that this paper was submitted.

\end{acknowledgements}

\bibliographystyle{aa} % style aa.bst
\bibliography{aa56000-25.bib}

\end{document}